\documentclass[onecolumn,aps,prd,preprintnumbers,showpacs,superscriptaddress,nofootinbib,floatfix,showkeys,notitlepage,longbibliography]{revtex4-2}

\usepackage[utf8]{inputenc} 
\usepackage[T1]{fontenc}
\usepackage{lmodern}
\usepackage{microtype}      

\usepackage{amsmath, amssymb, amsfonts, latexsym}
\usepackage{physics}
\usepackage{bm}             
\allowdisplaybreaks[4]      

\usepackage{graphicx}       
\usepackage{epstopdf}
\usepackage{float}
\usepackage{caption}
\usepackage{subcaption}     

\usepackage{booktabs}       
\usepackage{multirow}
\usepackage{array}
\usepackage{dcolumn}        
\usepackage{siunitx}        

\usepackage[normalem]{ulem} 
\usepackage{xcolor}         
\usepackage{comment}
\usepackage{blindtext}

\newcommand{\be}{\begin{equation}}
\newcommand{\ee}{\end{equation}}
\newcommand{\bq}{\begin{eqnarray}}
\newcommand{\eq}{\end{eqnarray}}

\usepackage{hyperref}
\hypersetup{
    colorlinks=true,
    linkcolor=blue,
    urlcolor=blue,
    citecolor=blue
}
\usepackage{orcidlink}      

\newcommand{\wR}{\omega_R}                
\newcommand{\Veff}{V_{ eff}}           

\newcommand{\fM}{\mathcal{F}}
\newcommand{\Dred}{\mathcal{D}}

\begin{document}

\title{Breaking Parameter Degeneracies in a Magnetically Charged Black Hole Embedded in a Hernquist Dark-Matter Halo: A Multi-Observable Analysis}

\author{Ali \"Ovg\"un \orcidlink{0000-0002-9889-342X}}
\email{ali.ovgun@emu.edu.tr} 
\affiliation{Physics Department, Faculty of Arts and Sciences Eastern Mediterranean University, Famagusta 99628 North Cyprus via Mersin 10, Turkiye.}

\author{Reggie C. Pantig \orcidlink{0000-0002-3101-8591}}
\email{rcpantig@mapua.edu.ph}
\affiliation{Physics Department, School of Foundational Studies and Education Map\'ua University, 658 Muralla St., Intramuros, Manila 1002, Philippines.}

\author{Joel Saavedra \orcidlink{0000-0002-1430-3008}}
\email{joel.saavedra@pucv.cl}
\affiliation{Instituto de F\'{\i}sica, Pontificia Universidad Cat\'olica de Valpara\'{\i}%
so, Casilla 4950, Valpara\'{\i}so, Chile.}

\date{\today}

\begin{abstract}
We study the intrinsic-environmental degeneracies for the
black-hole observables in a static spacetime sourced by a
nonlinear magnetic monopole immersed in a Hernquist dark-matter
halo. The geometry provides an analytic framework in which compact
strong-field modifications and extended environmental effects
enter the lapse function with opposite signs, yielding partially
compensating signatures in observables. We consider four complementary probes: the shadow radius $R_{ sh}$, eikonal quasinormal-mode (QNM) frequencies
$M\omega_R$, weak gravitational lensing $\hat{\theta}_\infty$,
and neutrino-antineutrino annihilation
$\dot{Q}/\dot{Q}_{ Newt}$, and map their degeneracy contours
in the $(g/M,\,\alpha/M)$ parameter plane at fixed halo
concentration $\beta/M$.
We demonstrate that single-observable diagnostics are insufficient
to uniquely determine the magnetic charge and the halo amplitude,
as different parameter combinations yield signatures nearly
indistinguishable from those of a Schwarzschild black hole.
The degeneracy paths of different observables are, however,
mutually non-parallel: the slopes $d\alpha/dg$ along
constant-$R_{ sh}$ and constant-$M\omega_R$ contours differ
by a factor of approximately five, so that their combination
breaks the remaining degeneracy and simultaneously constrains
both the intrinsic NED charge and the environmental dark-matter
distribution.
We derive the scalar, electromagnetic, and effective axial gravitational
perturbation equations and compute the quasinormal spectra using
a high-order WKB approach supplemented by Pad\'e resummation.
The magnetic charge raises the real oscillation frequency while
the Hernquist halo lowers it; for representative parameters the
two effects nearly cancel, but the cancellation is
observable-dependent and does not persist across all four
channels simultaneously.
For the shadow, we develop an expansion around an asymptotically
renormalized Schwarzschild background of mass
$\mathcal{M}=M+\alpha$, showing that at fixed $\mathcal{M}$
both sectors reduce $R_{sh}$ at first perturbative order.
For weak lensing, the leading deflection is determined by
$\mathcal{M}$ alone, while the first subleading correction
depends on $\mathcal{Q}=g^2+4\alpha\beta$, separating total
halo mass from halo concentration.
For neutrino-pair annihilation, the magnetic charge suppresses
the deposition rate by raising the lapse, while the halo
enhances it through the opposite sign, providing a third
independent constraint. Our results demonstrate that ringdown, imaging, lensing, and
high-energy deposition all probe the NED-halo competition with
different radial weightings, and that a combined multi-observable
analysis can resolve parameter degeneracies that any single
channel leaves unbroken.
\end{abstract}

\keywords{Quasinormal modes; Nonlinear electrodynamics;
Dark matter halo; Hernquist profile; Black hole;
WKB approximation; Neutrino annihilation.}

\maketitle


\section{Introduction}
\label{sec:intro}
The direct detection of gravitational waves from compact-binary
coalescences by the LIGO-Virgo-KAGRA network, and the
horizon-scale images of M87$^*$ and Sgr\,A$^*$ by the Event
Horizon Telescope mark the starting point in a precision observational era
of black-hole physics~\cite{Abbott2016,Abbott2021,EventHorizonTelescope:2019dse,
EventHorizonTelescope:2022wkp}.
In this context two observables are of particular interest.
The first is the ringdown signal, with its damped oscillations
described by quasinormal modes (QNMs)\cite{Kokkotas1999,Daghigh:2024wcl,Daghigh:2023ixh,Daghigh:2021psm}.
The second is the optical image in the strong field, encoded in
the photon ring and shadow.
These probes are complementary,  as they depend on related but not identical features of the
underlying geometry.

The quasinormal modes are complex frequencies set by the background
spacetime, the spin of the perturbing field, and the boundary
conditions of purely ingoing waves at the horizon and purely
outgoing waves at
infinity~\cite{Berti2009,Konoplya2011}.
They are therefore central to black-hole spectroscopy and tests of
the no-hair paradigm~\cite{Dreyer2004,Berti:2025hly}.
In the eikonal regime, the real part of the QNM frequency and the
damping rate are determined by the angular frequency and Lyapunov
exponent of the unstable photon orbit~\cite{Cardoso:2008bp},
directly connecting ringdown observables with shadow measurements
and motivating a unified treatment of wave dynamics and null
geodesics.

Most idealized studies assume an isolated vacuum black hole.
Astrophysical black holes, however, sit in galactic environments
whose dark-matter distribution may affect geodesics, redshift
factors, and wave propagation~\cite{Konoplya:2021ube,
Pezzella:2024tkf,Cardoso:2021wlq, Konoplya:2022hbl}. The Hernquist profile~\cite{Hernquist1990} is particularly suitable for this purpose because it has a finite total mass and an analytic
enclosed-mass function, while capturing a centrally concentrated
halo structure. It therefore provides a tractable model for studying the
modifications of strong-gravity observables by a realistic extended
matter distribution.

A second, independent deformation can arise from the compact object
itself. Nonlinear electrodynamics (NED) generalizes the Einstein--Maxwell
sector in the strong-field regime and has been widely used to
construct magnetically charged and regular black-hole
solutions~\cite{AyonBeato1998,Bronnikov2001,Bronnikov2022, Konoplya:2025ect}.
In such models the NED sector can modify the causal structure, the
photon sphere, the effective potential barrier governing
perturbations, and the corresponding ringdown spectrum.
Magnetically charged NED black holes are therefore useful
phenomenological laboratories for controlled departures from the
Reissner-Nordstr\"om paradigm.

The geometry studied in this paper combines both
ingredients~\cite{Jha2025NED, Jha2025Hernquist, Feng:2025iao}, a magnetically charged NED black
hole embedded in a Hernquist dark-matter halo.
The lapse contains a positive magnetic contribution and a negative
halo contribution, so the two sectors compete naturally.
Earlier analyses of this spacetime showed that strong-lensing and
optical observables can display partial cancellations between the
NED and halo parameters~\cite{Jha2025Hernquist, Feng:2025iao}.
This raises a sharper question for black-hole phenomenology:
does the same cancellation occur in the QNM spectrum, and can
ringdown, shadow, weak-lensing, and high-energy observables break
the resulting degeneracy?

Black-hole shadows have become a powerful probe of strong-field
gravity, since the size and shape of the photon-ring image are
controlled by the near-horizon optical geometry.
Extensive studies have shown that shadow observables are sensitive
to rotation, charge, plasma effects, higher-dimensional corrections,
and deviations from Kerr
geometry~\cite{Vagnozzi:2022moj,Allahyari:2019jqz,Afrin:2022ztr}.
In this work we analyze the shadow using the recently developed
perturbative framework for gravitational shadows in static
spherically symmetric spacetimes~\cite{Kobialko:2024zhc,
Pantig:2025deu}, which separates the trivial ADM-mass shift from
genuine non-Schwarzschild corrections.

Weak gravitational lensing offers a complementary probe in the
large-distance regime.
The application of the Gauss-Bonnet theorem to gravitational
lensing~\cite{Gibbons:2008rj}, extended to finite-distance
configurations by Ref.~\cite{Ishihara:2016vdc},
provides a geometrical interpretation of the deflection angle in
terms of the optical metric curvature.
In this work we compute the weak deflection angle using the
reference-renormalized curvature-primitive Gauss-Bonnet formalism
of Ref.~\cite{Pantig:2026xjj}, which isolates the leading
asymptotic mass contribution from genuine short-range corrections.

Neutrino-antineutrino annihilation $\nu\bar\nu\to e^-e^+$
provides a further diagnostic.
It converts energy from a neutrino bath into an
electron-positron plasma and is relevant for compact-object
environments such as collapsing stars, hyperaccreting disks, and
gamma-ray-burst central
engines~\cite{Salmonson:1999es,Asano:2000ib,Asano:2000dq}.
The rate is determined by gravitational redshift, null-ray bending,
and the proper-volume element, making it a sensitive probe of the
same lapse function that controls QNMs and optical
observables~\cite{Lambiase:2020iul,Pantig:2025eda}.
The neutrino channel therefore provides an independent test of the
NED-halo competition.

In this paper we give a unified analysis of the magnetically charged
NED-Hernquist black hole, with emphasis on the parameter
degeneracy structure.
We derive the scalar, electromagnetic, and axial gravitational QNMs
using a high-order WKB method with Pad\'e
resummation~\cite{Iyer1987,Konoplya2003,Matyjasek:2026},
and map four complementary observables simultaneously onto the
$(g/M,\,\alpha/M)$ parameter plane.
We show that the degeneracy contours are mutually non-parallel,
with slopes $d\alpha/dg$ differing by a factor of approximately
five between the shadow and the eikonal QNM channel, so that their
combination breaks the degeneracy and simultaneously constrains
both parameters.
We also provide an independent verification that the
metric of Ref.~\cite{Jha2025NED} satisfies the Einstein field
equations with the correct source terms
(Appendix~\ref{app:verification}).

The paper is structured as follows.
The NED-Hernquist geometry is described in
Section~\ref{sec:metric}.
Section~\ref{sec:shadow} studies the shadow at fixed asymptotic
mass.
Section~\ref{sec:deflection} derives the weak deflection angle.
Section~\ref{sec:perturbations} presents the perturbation equations
and the WKB method,
and  discusses the QNM spectrum and its
connection with the shadow.
Section~\ref{sec:neutrino} develops the neutrino-annihilation
formalism. Section~\ref{sec:degeneracy} discusses parameter degeneracy and multi-observable constraints.
Section~\ref{sec:conclusions} summarizes our results.
We use units $G=c=\hbar=1$.
\section{The NED-Hernquist black-hole spacetime}
\label{sec:metric}

We begin by reviewing the geometry and by fixing the notation used throughout the paper.  The model combines two physically distinct ingredients.  The first is an intrinsic magnetic deformation generated by a nonlinear-electrodynamics sector.  The second is an external matter contribution described by a Hernquist halo.  This separation is useful because the magnetic charge mainly affects the near-horizon and photon-sphere region, whereas the halo controls the large-scale gravitational potential and the asymptotic mass.  The spacetime therefore provides a clean setting in which compact-object physics and environmental effects can be compared within the same exact metric.

The Hernquist density profile is \cite{Hernquist1990}
\begin{equation}
  \rho_H(r)=\frac{\rho_s}{\displaystyle \frac{r}{r_s}\left(1+\frac{r}{r_s}\right)^3},
  \label{eq:hernquist_density}
\end{equation}
where $\rho_s$ and $r_s$ are the characteristic density and scale radius.  The profile behaves as $\rho_H\sim r^{-1}$ near the center and as $\rho_H\sim r^{-4}$ at large radius.  Thus the distribution is centrally concentrated but has a finite total mass.  The enclosed mass is
\begin{equation}
  M_H(r)=\frac{2\pi\rho_s r_s^3 r^2}{(r+r_s)^2}
        \equiv \frac{\alpha r^2}{(r+\beta)^2},
  \label{eq:halo_mass}
\end{equation}
with $\alpha\equiv 2\pi\rho_s r_s^3$ and $\beta\equiv r_s$.  Near the center, $M_H(r)\sim (\alpha/\beta^2)r^2$, while $M_H(r)\to\alpha$ as $r\to\infty$.  Hence $\alpha$ sets the total halo mass scale and $\beta$ controls the halo concentration.

The magnetic sector is described by the NED Lagrangian~\cite{Jha2025NED}
\begin{equation}
\mathcal{L}(\mathcal{F})=-\frac{\mathcal{F}}{
\left(1+\sqrt{|\mathcal{F}|/(2g^2)}\right)^2},
\label{eq:ned_lagrangian}
\end{equation}
where $\mathcal{F}=F_{\mu\nu}F^{\mu\nu}/4$ and $g$ is the magnetic charge.  The weak-field limit is $\mathcal{L}\to-\mathcal{F}$, so Maxwell theory is recovered far from the source.  For a static magnetic monopole,
\begin{equation}
  F_{\theta\phi}=g\sin\theta,
  \qquad
  \mathcal{F}=\frac{g^2}{2r^4}.
\end{equation}
Thus the NED correction is strongest in the compact region close to the hole and decays rapidly outward.  This behavior contrasts with the halo contribution, which is controlled by the extended scale $\beta$.

Solving the Einstein equations with the NED stress-energy tensor and the anisotropic Hernquist fluid gives the static line element~\cite{Jha2025NED}
\begin{equation}
  ds^2=-f(r)dt^2+\frac{dr^2}{f(r)}+r^2d\Omega^2,
  \label{eq:metric}
\end{equation}
where
\begin{equation}
  f(r)=1-\frac{2M}{r}+\frac{g^2}{r(r+g)}-\frac{2\alpha r}{(r+\beta)^2}.
  \label{eq:lapse}
\end{equation}
The Schwarzschild limit is recovered for $g=\alpha=0$, while $g=0$ gives the Schwarzschild-Hernquist case.  The sign structure of Eq.~\eqref{eq:lapse} is central: the NED term raises the lapse relative to Schwarzschild, whereas the halo term lowers it.  As a result, the two sectors tend to shift redshift-sensitive and geodesic observables in opposite directions.

At large radius,
\begin{equation}
  f(r)=1-\frac{2(M+\alpha)}{r}+\frac{g^2+4\alpha\beta}{r^2}+\mathcal{O}(r^{-3}).
  \label{eq:lapse_asymptotic}
\end{equation}
The ADM mass is therefore $\mathcal{M}=M+\alpha$.  The magnetic charge first appears in the $1/r^2$ sector, confirming that it is a short-range deformation, while the halo contributes already to the $1/r$ mass monopole.  Near the black hole, however, both sectors can influence the horizon, the photon sphere, and the effective potential barrier.  Consequently, neither sector should be regarded as observationally irrelevant a priori; their relative importance depends on the radial region probed by a given observable.

\section{Shadow analysis}
\label{sec:shadow}

The black-hole shadow provides a direct probe of the strong-field region, since its size is determined by unstable circular photon orbits \cite{Abdujabbarov:2016hnw,Atamurotov:2013sca,Atamurotov:2015nra,Abdujabbarov:2009az,Tsukamoto:2014tja,Tsukamoto:2017fxq,Battista:2026nsx,Wang:2025fmz,Kumar:2020hgm,Kuang:2022xjp,Kuang:2022ojj}. In the present spacetime, the halo modifies the asymptotic gravitational mass through ($\mathcal{M}=M+\alpha$), so the comparison with Schwarzschild must be made carefully. While fixing the bare mass (M) may suggest an enlarged shadow due to the halo contribution, fixing the observable asymptotic mass ($\mathcal{M}$) isolates the genuine short-range corrections from the NED-halo structure. This latter viewpoint is more physically relevant for distant observers.

We now derive the shadow of the NED-Hernquist black hole by adapting the first-order perturbative framework of shadow spectroscopy to the static line element introduced in \cite{Kobialko:2024zhc,Vertogradov:2024dpa,Pantig:2025deu}. Since
\begin{equation}
f(r)=1-\frac{2(M+\alpha)}{r}+\frac{g^2+4\alpha\beta}{r^2}+\mathcal{O}(r^{-3}),
\label{3.1}
\end{equation}
the mass measured by an asymptotic observer is not the bare black-hole mass $M$ alone, but the total asymptotic mass
\begin{equation}
\mathcal{M}\equiv M+\alpha.
\label{3.2}
\end{equation}

Since the large-$r$ expansion~\eqref{3.1} has the same
form as the Reissner--Nordstr\"om metric with mass
$\mathcal{M} = M+\alpha$ and effective charge squared
$\mathcal{Q} = g^2 + 4\alpha\beta$, one might consider organizing
the perturbation theory around a Reissner--Nordstr\"om reference
with those parameters rather than around a Schwarzschild background. We prefer the Schwarzschild reference due to the combination $\mathcal{Q} = g^2 + 4\alpha\beta$ mixes
contributions of fundamentally different physical origin: the compact NED charge $g$, which affects primarily the near-horizon region, and the halo concentration $\beta$, which controls the large-scale mass distribution.
Absorbing $\mathcal{Q}$ into a Reissner--Nordstr\"om background
would obscure this separation and prevent a transparent identification of which sector drives each observable correction.
Second, the Reissner--Nordstr\"om photon sphere and shadow depend on both the mass and the charge of the reference geometry, introducing an additional parameter into the zeroth-order solution that complicates the perturbative bookkeeping without adding physical clarity.
The Schwarzschild reference, by contrast, has a single parameter
$\mathcal{M}$, leaves both the NED correction and the residual
halo contribution in the deformation function $h(r)$, and makes their competition transparent at first perturbative order.
The two expansions are of course consistent: the Schwarzschild
result expanded further in $\mathcal{Q}/\mathcal{M}^2 \ll 1$
reproduces the Reissner--Nordstr\"om result at the same order.
For shadow observables defined at infinity, it is therefore more consistent to organize the perturbation theory around a Schwarzschild reference with mass $\mathcal{M}$, rather than around a Schwarzschild geometry of mass $M$. This choice absorbs the entire $1/r$ contribution into the background and leaves only genuinely non-Schwarzschild corrections in the perturbation. In particular, we write
\begin{equation}
f(r)=f_0(r)+\delta\,h(r),
\qquad
f_0(r)=1-\frac{2\mathcal{M}}{r},
\label{3.3}
\end{equation}
where $\delta$ is a bookkeeping parameter set to unity at the end of the calculation, and
\begin{equation}
h(r)=\frac{g^2}{r(r+g)}+\frac{2\alpha\beta(2r+\beta)}{r(r+\beta)^2}.
\label{3.4}
\end{equation}
Equation \eqref{3.4} follows from a simple rearrangement of the exact lapse and makes transparent the role of the large-$r$ expansion: once the total asymptotic mass is fixed, both the magnetic NED sector and the residual Hernquist contribution appear as short-range deformations of the Schwarzschild background. For $g>0$, $\alpha>0$, and $\beta>0$, the function $h(r)$ is positive outside the horizon.

To keep the derivation close to the perturbative method used in the shadow-spectroscopy framework, we first consider a test particle of rest mass $m$ moving on the equatorial plane $\theta=\pi/2$. The corresponding Lagrangian is
\begin{equation}
2\mathcal{L}
=
-f(r)\dot t^{\,2}
+f(r)^{-1}\dot r^{\,2}
+r^2\dot\phi^{\,2},
\label{3.5}
\end{equation}
where the overdot denotes differentiation with respect to an affine parameter. Stationarity and axial symmetry imply the conserved energy $E$ and angular momentum $L$,
\begin{equation}
E=f(r)\dot t,
\qquad
L=r^2\dot\phi.
\label{3.6}
\end{equation}
Using the normalization condition $g_{\mu\nu}\dot x^\mu\dot x^\nu=-m^2$, we obtain the radial equation
\begin{equation}
\dot r^{\,2}
=
E^2
-f(r)\left(m^2+\frac{L^2}{r^2}\right).
\label{3.7}
\end{equation}
It is convenient to introduce the dimensionless quantities
\begin{equation}
\varepsilon\equiv \frac{m^2}{E^2},
\qquad
\ell\equiv \frac{L^2}{E^2},
\label{3.8}
\end{equation}
so that Eq. \eqref{3.7} becomes
\begin{equation}
\frac{\dot r^{\,2}}{E^2}
=
1-f(r)\left(\varepsilon+\frac{\ell}{r^2}\right).
\label{3.9}
\end{equation}
The unstable circular orbits that determine the shadow boundary satisfy the standard conditions $\dot r=0$ and $\partial_r(\dot r^{\,2})=0$. Solving these conditions for $\varepsilon$ and $\ell$, we find
\begin{equation}
\ell=\frac{r^3 f'(r)}{2f(r)^2},
\qquad
\varepsilon=\frac{1}{f(r)}\left(1-\frac{r f'(r)}{2f(r)}\right).
\label{3.10}
\end{equation}
These relations are exact for any static, spherically symmetric metric with areal radius $r$. The shadow seen by a static observer at infinity is obtained by projecting the critical orbit onto the observer's celestial screen. The resulting squared shadow radius is
\begin{equation}
R^2(\varepsilon)
=
\frac{r^2}{f(r)}
\frac{1-\varepsilon f(r)}{1-\varepsilon},
\label{3.11}
\end{equation}
where $r=r_c(\varepsilon)$ is the radius of the unstable circular orbit. Using Eq. \eqref{3.10}, Eq. \eqref{3.11} can be written in the equivalent form \cite{Pantig:2025deu}
\begin{equation}
R^2(\varepsilon)
=
\frac{r^3 f'(r)}{2f(r)^2(1-\varepsilon)}.
\label{3.12}
\end{equation}
The ordinary black-hole shadow corresponds to the photon limit $\varepsilon=0$. In that case Eq. \eqref{3.10} reduces to the familiar photon-sphere condition
\begin{equation}
r_{ ph}f'(r_{ ph})-2f(r_{ ph})=0,
\label{3.13}
\end{equation}
and Eq. \eqref{3.11} becomes
\begin{equation}
R_{ sh}^2
=
\frac{r_{ ph}^2}{f(r_{ ph})}
=
\frac{2r_{ ph}}{f'(r_{ ph})}.
\label{3.14}
\end{equation}
This equation deserves a brief physical interpretation,
the quantity $R_{sh}$ is the critical impact parameter
$b_{crit} = L/E\big|_{r=r_{ph}}$ of the marginally captured photon orbit. To see this, note that the radial turning-point condition $\dot r = 0$ at $r = r_{\rm ph}$ gives, from Eq.\eqref{3.7},
\begin{equation}
  b_{\rm crit}^2
    \equiv \frac{L^2}{E^2}\bigg|_{r_{\rm ph}}
    = \frac{r_{\rm ph}^2}{f(r_{\rm ph})}\,,
\end{equation}
which is precisely $R_{\rm sh}^2$.
For a static observer at spatial infinity, $b_{\rm crit}$
is directly observable as the angular radius of the dark disk
cast by the black hole against a bright background the
shadow.
The subscript ``sh'' therefore labels this quantity as the
shadow radius in units of $M$, and $R_{\rm sh} = 3\sqrt{3}\,M$
in the Schwarzschild limit.
Equation~\eqref{3.14} is exact for any static, spherically
symmetric metric with lapse $f(r)$; the perturbative expansion
that follows is introduced not because the exact formula is
unavailable, but because it separates the contribution of the
asymptotic mass $\mathcal{M}$ from the genuine non-Schwarzschild
corrections, as we explain below. One could of course evaluate Eq.~\eqref{3.14} numerically for any choice of $(g,\alpha,\beta)$ without further approximation.
The perturbative expansion we develop below serves a different and
complementary purpose: it provides an analytic decomposition that
isolates the physically distinct contributions of the two sectors.
To appreciate this, consider that the exact photon-sphere condition mixes the NED charge $g$ and the halo parameters $(\alpha,\beta)$ in a transcendental equation with no closed-form solution.
A numerical evaluation of $R_{sh}$ therefore gives a single
number that does not reveal whether a given shift originates from
the compact magnetic sector or from the extended dark-matter
environment.
The perturbative expansion around the asymptotically renormalized
Schwarzschild background of mass $\mathcal{M} = M + \alpha$ resolves this ambiguity: it produces a closed first-order formula,in which the NED contribution and
the residual Hernquist contribution appear as separate 
identifiable terms with distinct scaling in the parameters
$(g,\alpha,\beta)$. We now implement the first-order expansion around the asymptotically renormalized Schwarzschild background in Eq. \eqref{3.3}. Let \cite{Pantig:2025deu}
\begin{equation}
r_c(\varepsilon,\delta)
=
r_0(\varepsilon)+\delta\,r_1(\varepsilon)+\mathcal{O}(\delta^2).
\label{3.15}
\end{equation}
The zeroth-order orbit is obtained by inserting $f_0(r)=1-2\mathcal{M}/r$ into Eq. \eqref{3.10}, which yields
\begin{equation}
\varepsilon
=
\frac{r(r-3\mathcal{M})}{(r-2\mathcal{M})^2}.
\label{3.16}
\end{equation}
Solving Eq. \eqref{3.16} for the unstable branch gives
\begin{equation}
r_0(\varepsilon)
=
\mathcal{M}\,
\frac{3-4\varepsilon+\sqrt{9-8\varepsilon}}{2(1-\varepsilon)}.
\label{3.17}
\end{equation}
This branch interpolates continuously between the photon sphere and the marginally bound timelike orbit,
\begin{equation}
r_0(0)=3\mathcal{M},
\qquad
\lim_{\varepsilon\to 1^-}r_0(\varepsilon)=4\mathcal{M}.
\label{3.18}
\end{equation}
Substituting Eq. \eqref{3.17} into Eq. \eqref{3.11}, we obtain the Schwarzschild background shadow radius
\begin{equation}
R_0^2(\varepsilon)
=
\frac{r_0(\varepsilon)^3}{4\mathcal{M}-r_0(\varepsilon)}.
\label{3.19}
\end{equation}
Equation \eqref{3.19} reproduces the usual photon-shadow result $R_0(0)=3\sqrt{3}\,\mathcal{M}$ when $\varepsilon=0$.

To determine the first-order shift of the unstable orbit, we define \cite{Pantig:2025deu}
\begin{equation}
\mathcal{F}(r,\delta)
\equiv
\frac{1}{f(r,\delta)}
\left(
1-\frac{r f'(r,\delta)}{2f(r,\delta)}
\right),
\label{3.20}
\end{equation}
so that the circular-orbit condition is $\varepsilon=\mathcal{F}(r_c,\delta)$. Expanding Eq. \eqref{3.20} around $(r_0,0)$ and requiring the coefficient of $\delta$ to vanish, we obtain

\begin{equation}
\begin{split}
r_1(\varepsilon)
&=
\frac{r_0(\varepsilon)^2}{2\mathcal{M}\,[6\mathcal{M}-r_0(\varepsilon)]}
\Bigl[
2\bigl(4\mathcal{M}-r_0(\varepsilon)\bigr)h\bigl(r_0(\varepsilon)\bigr)
- r_0(\varepsilon)\bigl(r_0(\varepsilon)-2\mathcal{M}\bigr)
h'\bigl(r_0(\varepsilon)\bigr)
\Bigr].
\end{split}
\label{3.21}
\end{equation}
For completeness, the derivative of the deformation profile is
\begin{equation}
h'(r)
=
-\frac{g^2(2r+g)}{r^2(r+g)^2}
-\frac{2\alpha\beta(\beta^2+3\beta r+4r^2)}{r^2(r+\beta)^3}.
\label{3.22}
\end{equation}
Equation \eqref{3.21} determines how the magnetic charge and the halo displace the location of the unstable orbit. However, a central structural simplification of the first-order formalism is that the explicit first-order correction to the shadow size does not require the knowledge of $r_1(\varepsilon)$. To see this, we define
\begin{equation}
\mathcal{G}(r,\delta)
\equiv
\frac{r^2}{f(r,\delta)}
\frac{1-\varepsilon f(r,\delta)}{1-\varepsilon},
\label{3.23}
\end{equation}
so that $R^2(\varepsilon,\delta)=\mathcal{G}(r_c(\varepsilon,\delta),\delta)$. The total derivative of $\mathcal{G}$ with respect to $\delta$ contains a term proportional to $\partial_r\mathcal{G}$, but this term vanishes at the background orbit because the shadow functional is stationary there. Consequently, the first-order correction depends only on the explicit metric deformation evaluated at the zeroth-order orbit. After a straightforward calculation, we find
\begin{equation}
R^2(\varepsilon)
=
R_0^2(\varepsilon)
-\delta\,
\frac{r_0(\varepsilon)^4}{\mathcal{M}\,[4\mathcal{M}-r_0(\varepsilon)]}
\,h\bigl(r_0(\varepsilon)\bigr)
+\mathcal{O}(\delta^2).
\label{3.24}
\end{equation}
Using Eq. \eqref{3.19}, Eq. \eqref{3.24} can be rewritten in the compact multiplicative form
\begin{equation}
R^2(\varepsilon)
=
R_0^2(\varepsilon)
\left[
1-\delta\,\frac{r_0(\varepsilon)}{\mathcal{M}}\,
h\bigl(r_0(\varepsilon)\bigr)
\right]
+\mathcal{O}(\delta^2).
\label{3.25}
\end{equation}
Substituting Eq. \eqref{3.4} into Eq. \eqref{3.25}, we obtain the first-order shadow radius in terms of the physical parameters of the NED-Hernquist spacetime,
\begin{equation}
\begin{split}
R^2(\varepsilon)
&=
R_0^2(\varepsilon)
\Biggl[
1-\delta \Biggl(
\frac{g^2}{\mathcal{M}\,[r_0(\varepsilon)+g]}
+\frac{2\alpha\beta\,[2r_0(\varepsilon)+\beta]}
{\mathcal{M}\,[r_0(\varepsilon)+\beta]^2}
\Biggr)
\Biggr]
+\mathcal{O}(\delta^2).
\end{split}
\label{3.26}
\end{equation}
Equation \eqref{3.26} is the central first-order result of this section. It shows that once the asymptotic mass shift has been absorbed into the reference geometry, both the magnetic NED sector and the residual Hernquist sector decrease the shadow relative to a Schwarzschild black hole of the same total mass $\mathcal{M}$. This sign is not accidental. For positive parameters, Eq. \eqref{3.4} implies $h(r)>0$, and therefore the correction term in Eq. \eqref{3.26} is strictly negative. Thus, at fixed ADM mass, the NED-Hernquist shadow is always smaller than the Schwarzschild shadow at first order.

The ordinary photon shadow is obtained by setting $\varepsilon=0$ in Eqs. \eqref{3.17}, \eqref{3.19}, and \eqref{3.26}. Since $r_0(0)=3\mathcal{M}$, we find
\begin{equation}
R_{ sh}^2
=
27\mathcal{M}^2
\left[
1-3\delta\,h(3\mathcal{M})
\right]
+\mathcal{O}(\delta^2),
\label{3.27}
\end{equation}
or, equivalently, at the level of the radius itself,
\begin{equation}
R_{ sh}
=
3\sqrt{3}\,\mathcal{M}
\left[
1-\frac{3}{2}\delta\,h(3\mathcal{M})
\right]
+\mathcal{O}(\delta^2).
\label{3.28}
\end{equation}
Substituting Eq. \eqref{3.4} into Eq. \eqref{3.28} yields the explicit first-order shadow radius
\begin{equation}
\begin{split}
R_{ sh}
&=
3\sqrt{3}\,\mathcal{M}
\Biggl[
1
-\delta \Biggl(
\frac{g^2}{2\mathcal{M}(3\mathcal{M}+g)}
+
\frac{\alpha\beta(6\mathcal{M}+\beta)}
{\mathcal{M}(3\mathcal{M}+\beta)^2}
\Biggr)
\Biggr]
+\mathcal{O}(\delta^2).
\end{split}
\label{3.29}
\end{equation}
Equation~\eqref{3.29} shows that, at fixed asymptotic mass, the first-order correction to the photon-shadow radius is negative for positive $g$, $\alpha$, and $\beta$.  The magnetic contribution is controlled by the compact scale $g$ and begins quadratically for small charge, whereas the residual halo correction depends on the concentration combination $\alpha\beta(6\mathcal{M}+\beta)/(3\mathcal{M}+\beta)^2$.  This does not contradict the bare-mass expansion, in which the halo can increase the shadow through its contribution to the total mass.  Rather, the two expansions answer different physical questions.  The fixed-$M$ result compares different ADM masses, while the fixed-$\mathcal{M}$ result isolates genuine non-Schwarzschild structure at the same asymptotic mass.
\begin{equation}
\frac{g^2}{2\mathcal{M}(3\mathcal{M}+g)}
=
\frac{g^2}{6\mathcal{M}^2}
+\mathcal{O}\!\left(\frac{g^3}{\mathcal{M}^3}\right),
\label{3.30}
\end{equation}
so the magnetic contribution enters quadratically. By contrast, the halo contribution behaves as
\begin{equation}
\frac{\alpha\beta(6\mathcal{M}+\beta)}
{\mathcal{M}(3\mathcal{M}+\beta)^2}
=
\frac{2\alpha\beta}{3\mathcal{M}^2}
+\mathcal{O}\!\left(\frac{\alpha\beta^2}{\mathcal{M}^3}\right)
\qquad
(\beta\ll \mathcal{M}),
\label{3.31}
\end{equation}
while for a very extended halo it approaches
\begin{equation}
\frac{\alpha\beta(6\mathcal{M}+\beta)}
{\mathcal{M}(3\mathcal{M}+\beta)^2}
\to
\frac{\alpha}{\mathcal{M}}
\qquad
(\beta\gg \mathcal{M}).
\label{3.32}
\end{equation}
These limits are physically sensible. A weak magnetic charge produces only a mild compact correction, whereas a diffuse but finite-mass halo can still noticeably reduce the shadow once the total asymptotic mass is held fixed.

It is useful to compare Eq. \eqref{3.29} with the exact photon-sphere condition written directly from the original lapse without asymptotic mass renormalization. Substituting the full $f(r)$ into Eq. \eqref{3.13}, we obtain
\begin{equation}
-2+\frac{6M}{r_{ ph}}
-\frac{g^2(4r_{ ph}+3g)}{r_{ ph}(r_{ ph}+g)^2}
+\frac{2\alpha r_{ ph}(3r_{ ph}+\beta)}{(r_{ ph}+\beta)^3}
=0.
\label{3.33}
\end{equation}
Equation \eqref{3.33} shows that if one insists on comparing the geometry to a Schwarzschild black hole of fixed bare mass $M$, the magnetic and halo contributions enter with opposite signs at the level of the photon-sphere equation. This leads to the first-order bare-mass expansion
\begin{equation}
R_{ sh}
=
3\sqrt{3}\,M
\left[
1
-\frac{g^2}{2M(3M+g)}
+\frac{9\alpha M}{(3M+\beta)^2}
\right]
+\mathcal{O}(\delta^2).
\label{3.34}
\end{equation}
The halo term in Eq.~\eqref{3.34} adds to the shadow because in that 
parametrization it acts partly by increasing the total mass seen at infinity. 
The two formulas address different physical questions: Eq.~\eqref{3.34} 
compares the NED-Hernquist geometry to a Schwarzschild spacetime of the same 
bare black-hole mass $M$, while Eq.~\eqref{3.29} compares it to a Schwarzschild 
spacetime of the same asymptotic mass $\mathcal{M}=M+\alpha$. Since the 
large-$r$ expansion in Eq.~\eqref{3.1} shows that the far-field geometry is 
controlled by $\mathcal{M}$, we regard Eq.~\eqref{3.29} as the more appropriate 
description of the observable shadow for a distant observer.

Two mechanisms govern the shadow of the NED-Hernquist black hole. The magnetic 
NED term modifies the compact near-hole structure and suppresses the shadow via 
the dimensionless ratio $g^2/[\mathcal{M}(3\mathcal{M}+g)]$. The Hernquist 
environment alters the mass distribution at larger scales and contributes through 
the combination $\alpha\beta(6\mathcal{M}+\beta)/[\mathcal{M}(3\mathcal{M}+\beta)^2]$. 
At fixed asymptotic mass both effects reduce the shadow at first order, reflecting 
the fact that the halo shifts the $1/r$ sector of the lapse while the NED 
correction remains localized in the shorter-range terms. The perturbative 
expansion is reliable as long as the dimensionless combination multiplying 
$\delta$ in Eq.~\eqref{3.29} is much smaller than unity, which is precisely the 
regime in which the exact NED-Hernquist geometry remains close to its 
asymptotically equivalent Schwarzschild background near the unstable circular 
orbit.

\begin{figure}
    \centering
    \includegraphics[width=0.8\textwidth]{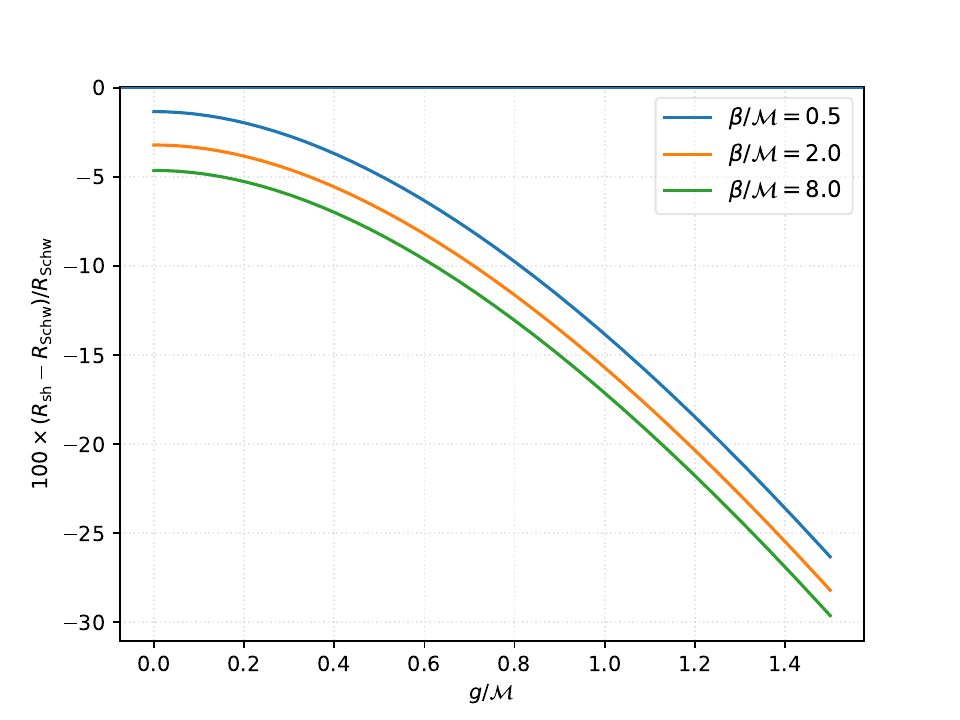}
    \caption{First-order fractional photon-shadow shift relative to a 
    Schwarzschild black hole with the same asymptotic mass \(\mathcal{M}\), 
    plotted against the magnetic charge parameter \(g/\mathcal{M}\). The three 
    curves correspond to different halo concentration scales \(\beta/\mathcal{M}\) 
    at fixed \(\alpha/\mathcal{M}=0.05\).}
    \label{fig1}
\end{figure}

Figure~\ref{fig1} shows the first-order fractional photon-shadow shift relative 
to a Schwarzschild black hole of the same asymptotic mass $\mathcal{M}$. All 
curves are negative, confirming that both residual corrections reduce the shadow 
at fixed ADM mass. Increasing $g/\mathcal{M}$ produces a monotonic decrease 
driven by the compact magnetic sector, while varying $\beta/\mathcal{M}$ shifts 
the offset through the concentration-dependent halo term.

\section{Weak deflection angle analysis}
\label{sec:deflection}

We next analyze the weak deflection of light in the same geometry.  Unlike the shadow, which is controlled by the photon-sphere region, weak lensing probes the outer optical geometry and is therefore especially sensitive to the asymptotic structure of the lapse.  Since the spacetime is asymptotically flat and $r$ is the areal radius, the natural reference optical geometry is Minkowski space \cite{Pantig:2026xjj}.  We use the reference-renormalized curvature-primitive form of the Gauss-Bonnet method, which expresses the optical curvature contribution as a radial boundary primitive and fixes the additive constant by requiring the discrepancy to vanish at infinity.

 We therefore work directly with the static, spherically symmetric line element
\begin{equation}
ds^2=-f(r)\,dt^2+f(r)^{-1}dr^2+r^2d\Omega^2,
\label{4.1}
\end{equation}
where
\begin{equation}
f(r)=1-\frac{2M}{r}+\frac{g^2}{r(r+g)}-\frac{2\alpha r}{(r+\beta)^2}.
\label{4.2}
\end{equation}
For null rays restricted to the equatorial plane, the optical metric is
\begin{equation}
d\sigma^2=\frac{dr^2}{f(r)^2}+\frac{r^2}{f(r)}\,d\phi^2.
\label{4.3}
\end{equation}
The first step in the reference-renormalized method is to express the Gaussian-curvature density as a total radial derivative. For a two-dimensional optical metric of the form in Eq. \eqref{4.3}, one finds
\begin{equation}
\mathcal{D}(r)\equiv K(r)\sqrt{\det \bar g(r)}
=
-\frac{d}{dr}\left[\sqrt{f(r)}-\frac{r f'(r)}{2\sqrt{f(r)}}\right].
\label{4.4}
\end{equation}
Hence the unique renormalized discrepancy primitive, normalized by the asymptotic condition $\mathcal{P}_{e}(\infty)=0$, is \cite{Pantig:2026xjj}
\begin{equation}
\mathcal{P}_{e}(r)
=
1-\sqrt{f(r)}+\frac{r f'(r)}{2\sqrt{f(r)}}.
\label{4.5}
\end{equation}
For the lapse in Eq. \eqref{4.2}, the derivative term is
\begin{equation}
\frac{r f'(r)}{2}
=
\frac{M}{r}
-\frac{g^{2}(2r+g)}{2r(r+g)^{2}}
+\frac{\alpha r(r-\beta)}{(r+\beta)^{3}}.
\label{4.6}
\end{equation}
Substituting Eq. \eqref{4.6} into Eq. \eqref{4.5} gives the exact primitive
\begin{equation}
\mathcal{P}_{e}(r)
=
1-\sqrt{f(r)}
+
\frac{1}{\sqrt{f(r)}}
\left[
\frac{M}{r}
-\frac{g^{2}(2r+g)}{2r(r+g)^{2}}
+\frac{\alpha r(r-\beta)}{(r+\beta)^{3}}
\right].
\label{4.7}
\end{equation}
This expression is exact and fully encodes the weak-lensing contribution of the spacetime in the photon-sphere-free formulation. For explicit weak-deflection calculations, however, it is more useful to expand Eq. \eqref{4.7} in the outer regime. To that end, we introduce the asymptotic combinations
\begin{equation}
\mathcal{M}\equiv M+\alpha,
\qquad
\mathcal{Q}\equiv g^{2}+4\alpha\beta.
\label{4.8}
\end{equation}
The large-$r$ lapse then takes the form
\begin{equation}
f(r)=1-\frac{2\mathcal{M}}{r}+\frac{\mathcal{Q}}{r^{2}}+\mathcal{O}(r^{-3}),
\label{4.9}
\end{equation}
where the omitted $r^{-3}$ sector contains the genuinely shorter-range terms inherited from the exact NED and Hernquist structure. Expanding Eq. \eqref{4.7} consistently in the weak-field regime gives
\begin{equation}
\mathcal{P}_{e}(r)
=
\frac{2\mathcal{M}}{r}
-\frac{3\mathcal{Q}}{2r^{2}}
+\mathcal{O}\!\left(\frac{\mathcal{M}^{2}}{r^{2}},\frac{\mathcal{M}\mathcal{Q}}{r^{3}},\frac{g^{3}+\alpha\beta^{2}}{r^{3}}\right).
\label{4.10}
\end{equation}
Equation \eqref{4.10} is the key input for the weak-deflection analysis. It shows immediately that the two sectors enter in qualitatively different ways. The halo mass scale $\alpha$ shifts the effective $1/r$ mass term through $\mathcal{M}=M+\alpha$, whereas both the magnetic charge and the halo concentration enter the $1/r^{2}$ sector through the combination $\mathcal{Q}=g^{2}+4\alpha\beta$. Therefore, at the order displayed, weak lensing is governed by the same two effective parameters that already control the far-field expansion of the metric.

We now evaluate the finite-distance weak deflection angle. In the asymptotically flat reference-renormalized formalism, the deflection angle is obtained by integrating the renormalized primitive along the straight reference ray of the Minkowski optical geometry. Let the source and receiver be located at radii $r_{S}$ and $r_{R}$, and let $b$ be the impact parameter. In the weak-deflection regime, the reference trajectory is
\begin{equation}
r^{(0)}(\phi)=\frac{b}{\cos\phi},
\qquad
-\Phi_{S}\le \phi \le \Phi_{R},
\label{4.11}
\end{equation}
where the endpoint angles are determined purely by the Euclidean reference geometry: \cite{Pantig:2026xjj}
\begin{equation}
\Phi_{R}=\arccos\!\left(\frac{b}{r_{R}}\right),
\qquad
\Phi_{S}=\arccos\!\left(\frac{b}{r_{S}}\right).
\label{4.12}
\end{equation}
We also define
\begin{equation}
F_{R}\equiv \sin\Phi_{R}=\sqrt{1-\frac{b^{2}}{r_{R}^{2}}},
\qquad
F_{S}\equiv \sin\Phi_{S}=\sqrt{1-\frac{b^{2}}{r_{S}^{2}}}.
\label{4.13}
\end{equation}
The finite-distance kinematic conditions require $b\le r_{S}$ and $b\le r_{R}$, while the weak-field approximation requires $\mathcal{M}/b\ll 1$ and $\mathcal{Q}/b^{2}\ll 1$.

Using Eqs. \eqref{4.10} and \eqref{4.11}, the finite-distance weak deflection angle, which we denote by $\hat{\vartheta}(b;r_S,r_R)$ to avoid confusion with the halo parameter $\alpha$, becomes \cite{Pantig:2026xjj}
\begin{equation}
\hat{\vartheta}(b;r_S,r_R)
=
\int_{-\Phi_{S}}^{\Phi_{R}}\mathcal{P}_{e}\!\left(r^{(0)}(\phi)\right)d\phi.
\label{4.14}
\end{equation}
Substituting the weak-field expansion of the primitive gives
\begin{equation}
\begin{split}
\hat{\vartheta}(b;r_S,r_R)
&=
\frac{2\mathcal{M}}{b}
\int_{-\Phi_{S}}^{\Phi_{R}}\cos\phi\,d\phi
\\
&\quad
-\frac{3\mathcal{Q}}{2b^{2}}
\int_{-\Phi_{S}}^{\Phi_{R}}\cos^{2}\phi\,d\phi
+\mathcal{O}(b^{-3}).
\end{split}
\label{4.15}
\end{equation}
where the remainder symbol collects the neglected higher-order weak-field contributions. The elementary integrals are
\begin{equation}
\int_{-\Phi_{S}}^{\Phi_{R}}\cos\phi\,d\phi
=
\sin\Phi_{R}+\sin\Phi_{S},
\label{4.16}
\end{equation}
and
\begin{equation}
\int_{-\Phi_{S}}^{\Phi_{R}}\cos^{2}\phi\,d\phi
=
\frac{1}{2}\left[
\Phi_{R}+\Phi_{S}
+\sin\Phi_{R}\cos\Phi_{R}
+\sin\Phi_{S}\cos\Phi_{S}
\right].
\label{4.17}
\end{equation}
Inserting Eqs. \eqref{4.16} and \eqref{4.17} into Eq. \eqref{4.15}, and then using Eqs. \eqref{4.12} and \eqref{4.13}, we obtain the finite-distance weak-deflection angle in closed form:
\begin{equation}
\begin{split}
\hat{\vartheta}(b;r_S,r_R)
&=
\frac{2\mathcal{M}}{b}\left(F_{R}+F_{S}\right)
\\
&\quad
-\frac{3\mathcal{Q}}{4b^{2}}
\Biggl[
\arccos\!\left(\frac{b}{r_{R}}\right)
+\arccos\!\left(\frac{b}{r_{S}}\right)
\\
&\qquad
+\frac{b}{r_{R}}F_{R}
+\frac{b}{r_{S}}F_{S}
\Biggr]
+\mathcal{O}(b^{-3}).
\end{split}
\label{4.18}
\end{equation}
Restoring the original parameters, Eq. \eqref{4.18} becomes
\begin{equation}
\begin{split}
\hat{\vartheta}(b;r_S,r_R)
&=
\frac{2(M+\alpha)}{b}\left(F_{R}+F_{S}\right)
\\
&\quad
-\frac{3\left(g^{2}+4\alpha\beta\right)}{4b^{2}}
\Biggl[
\arccos\!\left(\frac{b}{r_{R}}\right)
+\arccos\!\left(\frac{b}{r_{S}}\right)
\\
&\qquad
+\frac{b}{r_{R}}F_{R}
+\frac{b}{r_{S}}F_{S}
\Biggr]
+\mathcal{O}(b^{-3}).
\end{split}
\label{4.19}
\end{equation}
Equation~\eqref{4.19} shows three relevant properties of the finite-distance deflection angle. First, the leading $1/b$ term depends only on the asymptotic mass $\mathcal{M}=M+\alpha$. The magnetic NED sector does not contribute at this order, because in the far-field expansion it begins only at order $1/r^{2}$. Second, the first short-range correction is negative and proportional to $\mathcal{Q}=g^{2}+4\alpha\beta$. Since $g^{2}>0$ and $\alpha\beta>0$, the NED and halo-concentration pieces act in the same direction at this order: both reduce the deflection relative to a pure Schwarzschild lens with the same asymptotic mass. Third, the halo enters in a split manner. Its total mass $\alpha$ enhances the leading Newtonian term through $\mathcal{M}$, while its concentration scale $\beta$ contributes only through the short-range combination $4\alpha\beta$ in the $1/b^{2}$ correction. This separation is physically useful, because it shows that finite-distance lensing can distinguish overall halo mass from halo concentration already in the weak-field regime.

The limiting cases of Eq. \eqref{4.19} are also instructive. Setting $g=0$ and $\alpha=0$, we recover
\begin{equation}
\hat{\vartheta}_{ Schw}(b;r_S,r_R)
=
\frac{2M}{b}\left(F_{R}+F_{S}\right)
+\mathcal{O}(b^{-2}),
\label{4.20}
\end{equation}
which is the standard Schwarzschild finite-distance result. Setting $\alpha=0$ but retaining $g\neq 0$, we obtain
\begin{equation}
\begin{split}
\hat{\vartheta}_{ NED}(b;r_S,r_R)
&=
\frac{2M}{b}\left(F_{R}+F_{S}\right)
\\
&\quad
-\frac{3g^{2}}{4b^{2}}
\Biggl[
\arccos\!\left(\frac{b}{r_{R}}\right)
+\arccos\!\left(\frac{b}{r_{S}}\right)
\\
&\qquad
+\frac{b}{r_{R}}F_{R}
+\frac{b}{r_{S}}F_{S}
\Biggr]
+\mathcal{O}(b^{-3}).
\end{split}
\label{4.21}
\end{equation}
which has the same structure as the Reissner-Nordstr\"om finite-distance formula at the order retained. Finally, setting $g=0$ but keeping the halo parameters, we find
\begin{equation}
\begin{split}
\hat{\vartheta}_{ halo}(b;r_S,r_R)
&=
\frac{2(M+\alpha)}{b}\left(F_{R}+F_{S}\right)
\\
&\quad
-\frac{3\alpha\beta}{b^{2}}
\Biggl[
\arccos\!\left(\frac{b}{r_{R}}\right)
+\arccos\!\left(\frac{b}{r_{S}}\right)
\\
&\qquad
+\frac{b}{r_{R}}F_{R}
+\frac{b}{r_{S}}F_{S}
\Biggr]
+\mathcal{O}(b^{-3}).
\end{split}
\label{4.22}
\end{equation}
This last expression makes explicit that the halo affects weak lensing in two inequivalent ways: through the total outer mass scale $\alpha$ in the leading term and through the concentration product $\alpha\beta$ in the next order.

We now consider the regime in which the source and receiver are very far from the lensing object. Expanding Eq. \eqref{4.19} for $b/r_{S}\ll 1$ and $b/r_{R}\ll 1$, we use
\begin{equation}
\begin{split}
F_{i}
&=
1-\frac{b^{2}}{2r_{i}^{2}}
+\mathcal{O}\!\left(\frac{b^{4}}{r_{i}^{4}}\right),
\\[4pt]
\arccos\!\left(\frac{b}{r_{i}}\right)
&=
\frac{\pi}{2}
-\frac{b}{r_{i}}
-\frac{b^{3}}{6r_{i}^{3}}
+\mathcal{O}\!\left(\frac{b^{5}}{r_{i}^{5}}\right).
\end{split}
\label{4.23}
\end{equation}
for $i=S,R$. Combining these expansions, we find
\begin{equation}
\arccos\!\left(\frac{b}{r_{i}}\right)+\frac{b}{r_{i}}F_{i}
=
\frac{\pi}{2}
-\frac{2b^{3}}{3r_{i}^{3}}
+\mathcal{O}\!\left(\frac{b^{5}}{r_{i}^{5}}\right).
\label{4.24}
\end{equation}
Substituting Eqs. \eqref{4.23} and \eqref{4.24} into Eq. \eqref{4.19} yields the large-distance expansion
\begin{equation}
\begin{split}
\hat{\vartheta}(b;r_S,r_R)
&=
\frac{4\mathcal{M}}{b}
-\frac{3\pi\mathcal{Q}}{4b^{2}}
\\
&\quad
-\mathcal{M}b\left(\frac{1}{r_{R}^{2}}+\frac{1}{r_{S}^{2}}\right)
+\frac{\mathcal{Q}b}{2}\left(\frac{1}{r_{R}^{3}}+\frac{1}{r_{S}^{3}}\right)
\\
&\quad
+\mathcal{O}\!\left(b^{-3},\frac{b^{3}}{r_{i}^{4}}\right).
\end{split}
\label{4.25}
\end{equation}
Equation \eqref{4.25} clarifies the structure of finite-distance corrections. The leading mass sector acquires endpoint corrections at order $r_{i}^{-2}$, whereas the first short-range correction controlled by $\mathcal{Q}$ acquires endpoint corrections only at order $r_{i}^{-3}$. Thus the magnetic and halo-concentration sectors are less sensitive to finite-distance placement than the leading mass term. This is precisely what one expects from the hierarchy of the lapse function: the $1/r$ gravitational potential dominates the far-zone kinematics, while the $1/r^{2}$ corrections are more localized.

In the strict asymptotic limit $r_{S},r_{R}\to\infty$, Eq. \eqref{4.25} reduces to
\begin{equation}
\hat{\vartheta}_{\infty}(b)
=
\frac{4\mathcal{M}}{b}
-\frac{3\pi\mathcal{Q}}{4b^{2}}
+\mathcal{O}(b^{-3}),
\label{4.26}
\end{equation}
or, equivalently,
\begin{equation}
\hat{\vartheta}_{\infty}(b)
=
\frac{4(M+\alpha)}{b}
-\frac{3\pi\left(g^{2}+4\alpha\beta\right)}{4b^{2}}
+\mathcal{O}(b^{-3}).
\label{4.27}
\end{equation}
The hierarchy of weak-lensing effects is illustrated in 
Equations~\eqref{4.19} and~\eqref{4.27}. The leading deflection probes only 
the asymptotic mass $\mathcal{M}=M+\alpha$, whereas the first subleading 
correction probes the combination $\mathcal{Q}=g^2+4\alpha\beta$, coupling 
the intrinsic magnetic charge to the product of halo mass and concentration 
scale. The halo therefore enters weak lensing in two inequivalent ways: via 
its total mass in the Newtonian term, and via its concentration in the 
post-Newtonian correction. The shadow is dominated by the strong-field region 
near the unstable photon orbit, whereas the weak deflection angle is governed 
by the outer optical geometry; the two observables can thus disentangle 
compact NED effects from environmental halo effects in complementary radial 
regimes.

\begin{figure}
    \centering
    \includegraphics[width=0.8\textwidth]{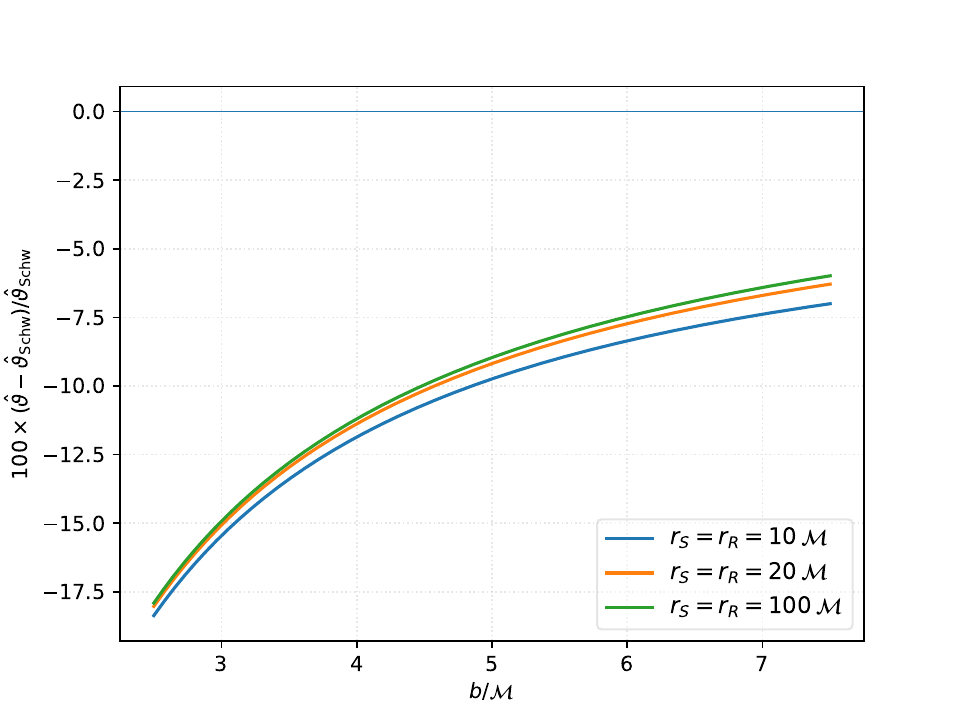}
    \caption{Fractional finite-distance weak-deflection shift, relative to a 
    Schwarzschild lens with the same asymptotic mass \(\mathcal{M}\), as a 
    function of impact parameter \(b/\mathcal{M}\). The three curves correspond 
    to symmetric source and receiver placements, \(r_S=r_R=10\mathcal{M}\), 
    \(20\mathcal{M}\) and \(100\mathcal{M}\), at fixed 
    \(\alpha/\mathcal{M}=0.05\), \(g/\mathcal{M}=0.6\) and 
    \(\beta/\mathcal{M}=2\).}
    \label{fig2}
\end{figure}

Figure~\ref{fig2} shows the fractional finite-distance deflection shift 
relative to Schwarzschild at the same asymptotic mass. The shift is negative 
since the comparison removes the common $1/b$ mass term, leaving the 
short-range correction proportional to $g^2+4\alpha\beta$. Its magnitude 
decreases with increasing impact parameter, following the $1/b^2$ scaling of 
the correction relative to the leading Schwarzschild term. The curve 
separation reflects finite-distance effects: the closer the source and 
receiver to the lens, the more sensitive the result is to endpoint geometry, 
and the asymptotic expression is recovered as $r_S,r_R\to\infty$.

\section{Perturbation equations and WKB method}
\label{sec:perturbations}

The dynamical response of the NED-Hernquist black hole is encoded in its quasinormal spectrum.  In the linear regime, perturbations reduce to wave propagation on the fixed background, and the geometry enters through an effective potential barrier.  This makes the QNM spectrum a sensitive probe of both the near-horizon magnetic deformation and the extended halo contribution.  In this section we write the master equations for scalar, electromagnetic, and axial gravitational perturbations and summarize the WKB method used in the numerical analysis.

The observational relevance of the present geometry ultimately rests on its dynamical response to small disturbances.
In the ringdown regime, the emitted signal is governed by a discrete set of damped oscillations, the quasinormal modes (QNMs), whose frequencies depend on the background spacetime and on the spin of the perturbing field.
For static and spherically symmetric black holes, the perturbation problem can often be reduced to a one-dimensional Schr\"odinger-type equation in the tortoise coordinate, which makes transparent how the geometry enters through an effective potential barrier \cite{Chandrasekhar1983,Kokkotas1999,Berti2009,Konoplya2011,Berti:2025hly}.
In the present NED-Hernquist background, this reduction is especially useful because it allows us to track the influence of the intrinsic magnetic deformation and the environmental halo contribution on wave propagation.

For a perturbing field with harmonic time dependence $e^{-i\omega t}$ and angular dependence described by spherical harmonics, the radial equation can be written as
\begin{equation}
  \frac{d^2\psi}{dr_*^2}+\left[\omega^2-V_{ eff}(r)\right]\psi=0,
  \label{eq:schrodinger}
\end{equation}
where the tortoise coordinate satisfies $dr_*/dr=f(r)^{-1}$.  The QNM boundary conditions are
\begin{equation}
  \psi\sim
  \begin{cases}
    e^{-i\omega r_*}, & r_*\to -\infty \quad (r\to r_h),\\[2mm]
    e^{+i\omega r_*}, & r_*\to +\infty \quad (r\to\infty),
  \end{cases}
  \label{eq:bc}
\end{equation}
which correspond to purely ingoing waves at the event horizon and purely outgoing waves at spatial infinity

The effective potentials used in this work are
\begin{align}
  V_{ eff}^{(0)}(r)&=f(r)\left[\frac{\ell(\ell+1)}{r^2}+\frac{f'(r)}{r}\right],
  \label{eq:V_scalar}\\
  V_{ eff}^{(1)}(r)&=f(r)\frac{\ell(\ell+1)}{r^2},
  \label{eq:V_em}\\
  V_{ eff}^{(2)}(r)&=f(r)\left[\frac{\ell(\ell+1)}{r^2}-\frac{f'(r)}{r}+\frac{2\bigl(f(r)-1\bigr)}{r^2}\right].
  \label{eq:Veff_s2}
\end{align}
The last expression reduces to the standard Regge-Wheeler potential $V_{ RW}=f[\ell(\ell+1)/r^2-6M/r^3]$ in the Schwarzschild limit.  We emphasize that for axial (odd-parity) gravitational perturbations of a static,
spherically symmetric spacetime, the perturbed Einstein equations
in the Regge-Wheeler gauge~\cite{Regge1957,Chandrasekhar1983} reduce to a single
master equation of the form~\eqref{eq:schrodinger} with the
potential $V_{\rm eff}^{(2)}$ given by Eq.~\eqref{eq:Veff_s2}.
The key point is that, for axial perturbations, the odd-parity
sector of the NED field equation becomes a constraint rather than a dynamical equation: the perturbed magnetic field
has no odd-parity degree of freedom that propagates
independently~\cite{Moreno:2002gg, Toshmatov:2018tyo}.
Consequently, the axial gravitational and NED perturbations
decouple at linear order, and Eq.~\eqref{eq:Veff_s2} captures the
complete dynamics of the odd-parity gravitational sector without
NED coupling corrections.
An analogous argument applies to the Hernquist fluid: for an
anisotropic perfect fluid in hydrostatic equilibrium, the
odd-parity fluid perturbations satisfy a constraint equation
at linear order and do not source the axial gravitational
modes~\cite{Chandrasekhar1983}.

The situation is different for even-parity (polar) perturbations,
which do mix gravitational, NED field, and fluid degrees of
freedom.
A fully coupled polar perturbation analysis of the
NED--Hernquist geometry would require treating the perturbed NED
Lagrangian and the perturbed anisotropic stress tensor
simultaneously, leading to a system of coupled
Zerilli-type equations.
This analysis is substantially more involved and is left for
future work.
The present study therefore provides the complete odd-parity
QNM spectrum and an effective Regge--Wheeler-type approximation
for the even-parity sector, which together give a reliable
characterization of the dominant ringdown modes with
$\ell \geq 2$.

We compute the QNM frequencies using the WKB approximation developed by Iyer and Will \cite{Iyer1987} and extended to higher orders by Konoplya and collaborators~\cite{Konoplya2003}. In the present work, we implement the WKB method up to
16th order, combined with Pad\'e resummation techniques~\cite{Matyjasek:2026},
which significantly improves the convergence and accuracy of the series. This approach
provides highly accurate results for the dominant modes with $\ell \geq 2$, including
both the real and imaginary parts of the quasinormal frequencies, as verified against Schwarzschild reference values (Table~\ref{tab:schwarzschild}).  The method approximates the effective potential near its peak
by a Taylor expansion and matches the asymptotic solutions,
yielding the quantization condition
\begin{equation}
  \frac{i(\omega^2-V_0)}{\sqrt{-2V_0''}}-\sum_{k=2}^{N}\Lambda_k
  =n+\frac{1}{2},
  \qquad N=16,
  \label{eq:wkb}
\end{equation}
where $V_0 \equiv V_{ eff}(r_0)$ is the potential at its maximum $r_0$,
$V_0'' \equiv d^2V_{ eff}/dr_*^2\big|_{r_0}$,
and $\Lambda_k$ are higher-order WKB correction terms that depend
on the derivatives of $V_{ eff}$ at $r_0$
(explicit expressions are given in Refs.~\cite{Konoplya2003,Iyer1987,Matyjasek:2026}).
The integer $n = 0, 1, 2, \ldots$ labels the overtone.  The method is most reliable for single-barrier potentials and for modes with $\ell\gtrsim n$, which is the regime considered below.  The Schwarzschild limit is used as a validation benchmark before varying the NED and halo parameters.

\begin{figure}[t]
  \centering
  \includegraphics[width=\textwidth]{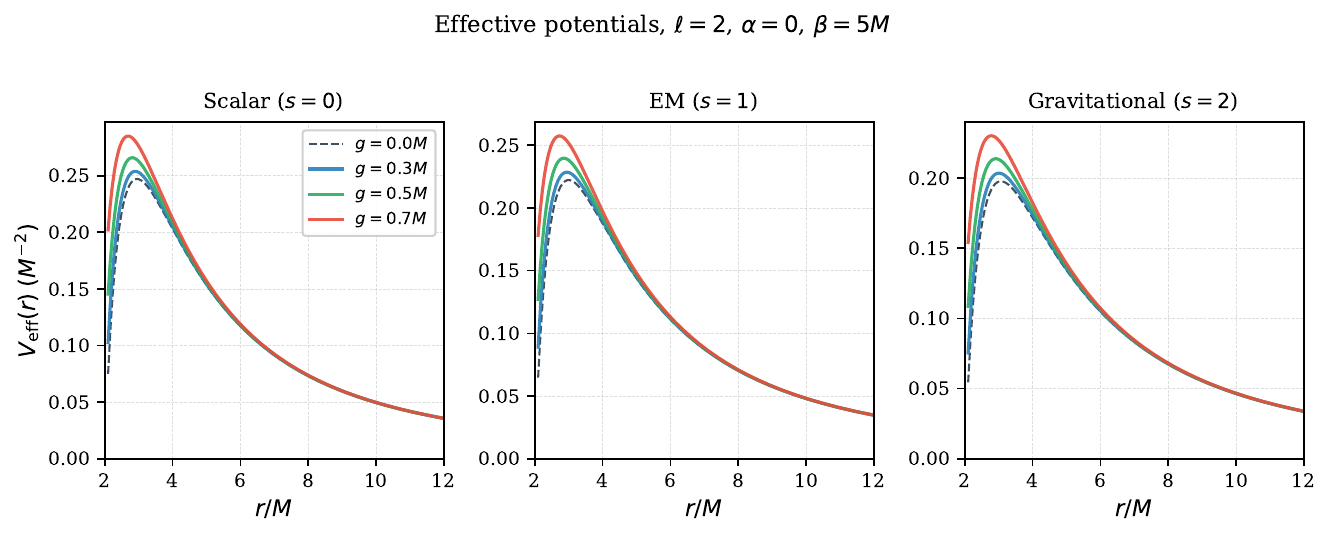}
  \caption{\parbox{0.95\textwidth}{Effective potentials $V_{ eff}(r)$ for scalar ($s=0$, left),
           electromagnetic ($s=1$, centre), and axial gravitational
           ($s=2$, right) perturbations with $\ell=2$, $\alpha=0$,
           $\beta=5M$, and several values of the magnetic charge $g$.
           The dashed curve ($g=0$) corresponds to the Schwarzschild limit.}}
  \label{fig:Veff}
\end{figure}

Figure~\ref{fig:Veff} shows the effective potential $\Veff(r)$ for
$\ell=2$ and several values of $g$, with $\alpha=0$ and $\beta=5M$.
Increasing $g$ raises the peak of the potential barrier and shifts it
slightly inward, consistent with the upward shift in $\wR$.

\subsection{Schwarzschild limit and method validation}
\label{subsec:validation}

Before presenting MHDM results, we validate the WKB implementation against the Schwarzschild case ($g=0$, $\alpha=0$).
Table~\ref{tab:schwarzschild} lists the computed frequencies for
$\ell=1$-$4$, $n=0$-$2$, and all three spins.
Specifically, for the dominant modes ($\ell \geq 2$, $n=0$), the agreement with reference values of ~\cite{Berti2009} confirms the reliability of the high-order WKB method with Padé resummation.

\begin{table}[t]
\caption{\label{tab:schwarzschild}
\textcolor{black}{Quasinormal-mode frequencies for the Schwarzschild black hole ($g=0$, $\alpha=0$, $\beta=5M$) in units $M=1$, computed using the WKB method up to 16th order with Pad\'e resummation. We write $\omega=\omega_R-i\omega_I$ with $\omega_I>0$.}}
\begin{ruledtabular}
\begin{tabular}{ccccc}
$s$ & $\ell$ & $n$ & $\omega_R$ & $\omega_I$ \\
\hline
\multicolumn{5}{c}{Scalar perturbations ($s=0$)} \\
\hline
0 & 1 & 0 & 0.292936 & 0.097660 \\
0 & 1 & 1 & 0.264448 & 0.306259 \\
0 & 2 & 0 & 0.483644 & 0.096759 \\
0 & 2 & 1 & 0.463851 & 0.295604 \\
0 & 2 & 2 & 0.430544 & 0.508558 \\
0 & 3 & 0 & 0.675366 & 0.096500 \\
0 & 3 & 1 & 0.660671 & 0.292285 \\
0 & 3 & 2 & 0.633626 & 0.496008 \\
0 & 4 & 0 & 0.867416 & 0.096392 \\
0 & 4 & 1 & 0.855808 & 0.290876 \\
0 & 4 & 2 & 0.833692 & 0.490325 \\
\hline
\multicolumn{5}{c}{Electromagnetic perturbations ($s=1$)} \\
\hline
1 & 1 & 0 & 0.248263 & 0.092488 \\
1 & 1 & 1 & 0.214510 & 0.293673 \\
1 & 2 & 0 & 0.457596 & 0.095004 \\
1 & 2 & 1 & 0.436542 & 0.290710 \\
1 & 2 & 2 & 0.401186 & 0.501586 \\
1 & 3 & 0 & 0.656899 & 0.095616 \\
1 & 3 & 1 & 0.641737 & 0.289728 \\
1 & 3 & 2 & 0.613832 & 0.492066 \\
1 & 4 & 0 & 0.853095 & 0.095860 \\
1 & 4 & 1 & 0.841267 & 0.289315 \\
1 & 4 & 2 & 0.818728 & 0.487838 \\
\hline
\multicolumn{5}{c}{Gravitational perturbations ($s=2$)} \\
\hline
2 & 2 & 0 & 0.373609 & 0.088976 \\
2 & 2 & 1 & 0.346584 & 0.273942 \\
2 & 3 & 0 & 0.599443 & 0.092703 \\
2 & 3 & 1 & 0.582644 & 0.281298 \\
2 & 3 & 2 & 0.551685 & 0.479093 \\
2 & 4 & 0 & 0.809178 & 0.094164 \\
2 & 4 & 1 & 0.796632 & 0.284334 \\
2 & 4 & 2 & 0.772710 & 0.479908 \\
\end{tabular}
\end{ruledtabular}
\end{table}

\textcolor{black}{Tables~\ref{tab:spin0_vs_g}-\ref{tab:spin2_vs_g} isolate the effect of the NED magnetic charge by setting $\alpha=0$.  The real part of the frequency increases monotonically with $g$ for all three perturbation sectors.  For the scalar fundamental mode with $\ell=2$, the shift is $+0.60\%$ at $g=0.2M$, $+5.33\%$ at $g=0.6M$, and $+9.70\%$ at $g=0.8M$.  The nearly spin-independent behavior of the fractional shift indicates that the main effect of $g$ is to raise the common potential barrier rather than to introduce a strongly spin-selective correction.  The imaginary part also increases mildly, corresponding to a slightly faster ringdown.}

\begin{table}[t]

\caption{\label{tab:spin0_vs_g}
\textcolor{black}{Fundamental quasinormal frequencies for scalar perturbations ($s=0$) as a function of $g/M$ for $\alpha=0$ and $\beta=5M$, in units $M=1$, computed using the WKB method up to 16th order with Pad\'e resummation. We write $\omega=\omega_R-i\omega_I$ with $\omega_I>0$. The last column shows the relative deviation in $\omega_R$ with respect to the Schwarzschild value.}}
\begin{ruledtabular}
\begin{tabular}{ccccccc}
$\ell$ & $g/M$ & $\omega_R$ & $\omega_I$ & $\omega_R^{Sch}$ & $\omega_I^{Sch}$ & $\Delta(\%)$ \\
\hline
\multicolumn{7}{c}{$\ell=1$} \\
\hline
1 & 0 & 0.292936 & 0.097660 & &  & 0.00  \\
1 & 0.2 & 0.294809 & 0.097889 & &  & 0.64 \\
1 & 0.4 & 0.300223 & 0.098649 & 0.292936 & 0.097660 & 2.49 \\
1 & 0.6 & 0.309277 & 0.100021 &  &  & 5.58 \\
1 & 0.8 & 0.322566 & 0.102122 &  & & 10.11 \\
\hline
\multicolumn{7}{c}{$\ell=2$} \\
\hline
2 & 0 & 0.483644 & 0.096759 & & & 0.00  \\
2 & 0.2 & 0.486718 & 0.096993 & & & 0.64 \\
2 & 0.4 & 0.495629 & 0.097762 & 0.483644 & 0.096759 & 2.48 \\
2 & 0.6 & 0.510559 & 0.099154 & & & 5.57 \\
2 & 0.8 & 0.532529 & 0.101282 & & &  10.11 \\
\hline
\multicolumn{7}{c}{$\ell=3$} \\
\hline
3 & 0 & 0.675366 & 0.096500 & && 0.00 \\
3 & 0.2 & 0.679651 & 0.096735 & & & 0.63 \\
3 & 0.4 & 0.692081 & 0.097507 & 0.675366 & 0.096500 & 2.47 \\
3 & 0.6 & 0.712921 & 0.098905 & && 5.56 \\
3 & 0.8 & 0.743608 & 0.101040 & && 10.10 \\
\hline
\multicolumn{7}{c}{$\ell=4$} \\
\hline
4 & 0 & 0.867416 & 0.096392 & & & 0.00  \\
4 & 0.2 & 0.872915 & 0.096627 & & & 0.63 \\
4 & 0.4 & 0.888872 & 0.097401 & 0.867416 & 0.096392 & 2.47 \\
4 & 0.6 & 0.915633 & 0.098801 & & & 5.56 \\
4 & 0.8 & 0.955049 & 0.100940 & & & 10.10 \\
\end{tabular}
\end{ruledtabular}
\end{table}

\begin{table}[t]

\caption{
\label{tab:spin1_vs_g}
\textcolor{black}{Fundamental quasinormal frequencies for electromagnetic perturbations ($s=1$) as a function of $g/M$ for $\alpha=0$ and $\beta=5M$, in units $M=1$, computed using the WKB method up to 16th order with Pad\'e resummation. We write $\omega=\omega_R-i\omega_I$ with $\omega_I>0$. The last column shows the relative deviation in $\omega_R$ with respect to the Schwarzschild value.}}
\begin{ruledtabular}
\begin{tabular}{ccccccc}
$\ell$ & $g/M$ & $\omega_R$ & $\omega_I$ & $\omega_R^{Sch}$ & $\omega_I^{Sch}$ & $\Delta(\%)$ \\
\hline
\multicolumn{7}{c}{$\ell=1$} \\
\hline
1 & 0 & 0.248263 & 0.092488 & & & 0.00\\
1 & 0.2 & 0.250040 & 0.092754 & & & 0.72 \\
1 & 0.4 & 0.255142 & 0.093603 & 0.248263 & 0.092488 & 2.77 \\
1 & 0.6 & 0.263633 & 0.095112 & & & 6.19 \\
1 & 0.8 & 0.276078 & 0.097388 & & & 11.20 \\
\hline
\multicolumn{7}{c}{$\ell=2$} \\
\hline
2 & 0 & 0.457596 & 0.095004 & & & 0.00 \\
2 & 0.2 & 0.460611 & 0.095249 & & & 0.66 \\
2 & 0.4 & 0.469333 & 0.096047 & 0.457596 & 0.095004 & 2.57 \\
2 & 0.6 & 0.483930 & 0.097483 & & & 5.76 \\
2 & 0.8 & 0.505403 & 0.099669 & && 10.45 \\
\hline
\multicolumn{7}{c}{$\ell=3$} \\
\hline
3 & 0 & 0.656899 & 0.095616 & & & 0.00\\
3 & 0.2 & 0.661141 & 0.095857 & & &0.65  \\
3 & 0.4 & 0.673436 & 0.096644 & 0.656899 & 0.095616 & 2.52 \\
3 & 0.6 & 0.694039 & 0.098063 & & & 5.65 \\
3 & 0.8 & 0.724372 & 0.100227 & & & 10.27 \\
\hline
\multicolumn{7}{c}{$\ell=4$} \\
\hline
4 & 0 & 0.853095 & 0.095860 & & & 0.00 \\
4 & 0.2 & 0.858562 & 0.096099 & & & 0.64 \\
4 & 0.4 & 0.874414 & 0.096881 & 0.853095 & 0.095860 & 2.50 \\
4 & 0.6 & 0.900990 & 0.098294 & & & 5.61 \\
4 & 0.8 & 0.940132 & 0.100450 & & & 10.20 \\
\end{tabular}
\end{ruledtabular}
\end{table}

\begin{table}[t]

\caption{\label{tab:spin2_vs_g}
\textcolor{black}{Fundamental quasinormal frequencies for axial gravitational perturbations ($s=2$) as a function of $g/M$ for $\alpha=0$ and $\beta=5M$, in units $M=1$, computed using the WKB method up to 16th order with Pad\'e resummation. We write $\omega=\omega_R-i\omega_I$ with $\omega_I>0$. The last column shows the relative deviation in $\omega_R$ with respect to the Schwarzschild value.}}

\begin{ruledtabular}
\begin{tabular}{ccccccc}
$\ell$ & $g/M$ & $\omega_R$ & $\omega_I$ & $\omega_R^{Sch}$ & $\omega_I^{Sch}$ & $\Delta(\%)$ \\
\hline
\multicolumn{7}{c}{$\ell=2$} \\
\hline
2 & 0 & 0.373609 & 0.088976 & & & 0.00 \\
2 & 0.2 & 0.376066 & 0.089207 && & 0.66 \\
2 & 0.4 & 0.383171 & 0.089965 & 0.373609 & 0.088976 & 2.56 \\
2 & 0.6 & 0.395079 & 0.091300 & & & 5.75 \\
2 & 0.8 & 0.412566 & 0.093411 & & & 10.43 \\
\hline
\multicolumn{7}{c}{$\ell=3$} \\
\hline
3 & 0 & 0.599443 & 0.092703 & & & 0.00 \\
3 & 0.2 & 0.603312 & 0.092944 & & & 0.65 \\
3 & 0.4 & 0.614524 & 0.093727 & 0.599443 & 0.092703 & 2.52 \\
3 & 0.6 & 0.633311 & 0.095137 & & & 5.65 \\
3 & 0.8 & 0.660973 & 0.097286 & & & 10.26 \\
\hline
\multicolumn{7}{c}{$\ell=4$} \\
\hline
4 & 0 & 0.809178 & 0.094164 & & & 0.00\\
4 & 0.2 & 0.814362 & 0.094404 && & 0.64 \\
4 & 0.4 & 0.829395 & 0.095187 & 0.809178 & 0.094164 & 2.50 \\
4 & 0.6 & 0.854598 & 0.096599 & & & 5.61 \\
4 & 0.8 & 0.891717 & 0.098752 & & & 10.20 \\
\end{tabular}
\end{ruledtabular}
\end{table}

\textcolor{black}{Table~\ref{tab:halo_effect} shows the complementary effect of the Hernquist halo at fixed $g=0.4M$ and $\beta=5M$.  Increasing $\alpha$ lowers $\omega_R$, thereby compensating the NED-induced increase.  For the scalar $\ell=2$ fundamental mode, the relative shift changes from a positive value at $\alpha=0$ to a small negative value at $\alpha M=0.20$, with a near-cancellation around $\alpha M\simeq0.15$.  At this point the MHDM spectrum becomes almost indistinguishable from Schwarzschild at the level of a single frequency.  This is the QNM counterpart of the degeneracy already observed in shadow and lensing observables, and it shows why a multi-observable analysis is necessary.}

\textcolor{black}{Tables II-IV show that, when the halo is switched off, the magnetic NED charge produces a monotonic increase in the real part of the quasinormal frequencies. By contrast, Table V shows that, at fixed $g=0.4M$, increasing the Hernquist parameter $\alpha$ decreases $\omega_R$, partially compensating the NED-induced enhancement. This leads to a near-degeneracy with the Schwarzschild spectrum around $\alpha M \simeq 0.15$ for the fundamental modes. Table VI confirms that this behavior persists for the first overtones.}
\begin{table*}[t] 
\caption{\label{tab:halo_effect}
Effect of the Hernquist dark-matter halo on quasinormal-mode frequencies. 
The parameters are fixed to $g=0.4M$, $\beta=5M$, and $n=0$, while the halo parameter $\alpha M$ varies from 0 to 0.20. Units are $M=1$. The last column shows the signed relative deviation $\Delta\omega_R = 100\,(\omega_R-\omega_R^{ Sch})/\omega_R^{ Sch}$, highlighting the competition between NED and halo effects.
}
\centering
\small
\setlength{\tabcolsep}{4pt}
\renewcommand{\arraystretch}{1.1}
\begin{ruledtabular}
\begin{tabular}{cccccc}
$\ell$ & $\alpha M$ & $\omega_R$ & $\omega_I$ & $\omega_R^{ Sch}$ & $\Delta(\%)$ \\
\hline

\multicolumn{6}{c}{Scalar perturbations ($s=0$)} \\
\hline
2 & 0.00 & 0.49563 & 0.09776 & 0.48364 & 2.48 \\
2 & 0.05 & 0.49205 & 0.09676 & 0.48364 & 1.74 \\
2 & 0.10 & 0.48847 & 0.09576 & 0.48364 & 1.00 \\
2 & 0.15 & 0.48490 & 0.09477 & 0.48364 & 0.26 \\
2 & 0.20 & 0.48133 & 0.09378 & 0.48364 & 0.48 \\

3 & 0.00 & 0.69208 & 0.09751 & 0.67537 & 2.47 \\
3 & 0.05 & 0.68713 & 0.09651 & 0.67537 & 1.74 \\
3 & 0.10 & 0.68219 & 0.09552 & 0.67537 & 1.01 \\
3 & 0.15 & 0.67724 & 0.09454 & 0.67537 & 0.28 \\
3 & 0.20 & 0.67230 & 0.09355 & 0.67537 & 0.45 \\

4 & 0.00 & 0.88887 & 0.09740 & 0.86742 & 2.47 \\
4 & 0.05 & 0.88254 & 0.09641 & 0.86742 & 1.74 \\
4 & 0.10 & 0.87621 & 0.09542 & 0.86742 & 1.01 \\
4 & 0.15 & 0.86989 & 0.09444 & 0.86742 & 0.28 \\
4 & 0.20 & 0.86356 & 0.09346 & 0.86742 & 0.44 \\

\hline
\multicolumn{6}{c}{Electromagnetic perturbations ($s=1$)} \\
\hline
2 & 0.00 & 0.46933 & 0.09605 & 0.45760 & 2.57 \\
2 & 0.05 & 0.46609 & 0.09508 & 0.45760 & 1.86 \\
2 & 0.10 & 0.46285 & 0.09411 & 0.45760 & 1.15 \\
2 & 0.15 & 0.45961 & 0.09315 & 0.45760 & 0.44 \\
2 & 0.20 & 0.45637 & 0.09219 & 0.45760 & 0.27 \\

\hline
\multicolumn{6}{c}{Gravitational perturbations ($s=2$)} \\
\hline
2 & 0.00 & 0.38317 & 0.08996 & 0.37361 & 2.56 \\
2 & 0.05 & 0.38050 & 0.08908 & 0.37361 & 1.84 \\
2 & 0.10 & 0.37783 & 0.08819 & 0.37361 & 1.13 \\
2 & 0.15 & 0.37517 & 0.08728 & 0.37361 & 0.42 \\
2 & 0.20 & 0.37250 & 0.08640 & 0.37361 & 0.30 \\

\end{tabular}
\end{ruledtabular}
\end{table*}

\textcolor{black}{Table~\ref{overtones} extends the comparison to overtones $n=0,1,2$ for the representative point $g=0.4M$, $\alpha=0.1M$, and $\beta=5M$.  The relative shifts remain at the percent level and vary smoothly with $n$.  This behavior indicates that the combined NED-halo deformation mainly changes the height and curvature of the potential barrier while preserving the overall structure of the spectrum.  From an observational perspective, this is useful: mode ratios are not dramatically reshuffled, but the coherent displacement of the spectrum can still encode the presence of the two sectors.}
\begin{table*}[t]
\caption{\label{overtones}
Quasinormal-mode frequencies including overtones ($n=0,1,2$)
for $g=0.4M$, $\alpha=0.1M$, $\beta=5M$ ($M=1$).
Schwarzschild reference corresponds to $g=0$, $\alpha=0$, units are $M=1$. The small relative deviations, typically at the percent level, indicate that the QNM spectrum remains close to the Schwarzschild case even in the presence of both NED and halo contributions.
}
\centering
\small
\setlength{\tabcolsep}{4pt}
\renewcommand{\arraystretch}{1.1}
\begin{ruledtabular}
\begin{tabular}{ccccccc}
$\ell$ & $n$ & $\omega_R$ & $\omega_I$ &
$\omega_R^{ Sch}$ & $\omega_I^{ Sch}$ & $\Delta(\%)$ \\
\hline

\multicolumn{7}{c}{Scalar perturbations ($s=0$)} \\
\hline
2 & 0 & 0.48847 & 0.09576 & 0.48364 & 0.09676 & 1.00 \\
2 & 1 & 0.46951 & 0.29233 & 0.46385 & 0.29560 & 1.22 \\
2 & 2 & 0.43744 & 0.50229 & 0.43054 & 0.50856 & 1.60 \\
3 & 0 & 0.68219 & 0.09552 & 0.67537 & 0.09650 & 1.01 \\
3 & 1 & 0.66812 & 0.28921 & 0.66067 & 0.29228 & 1.13 \\
3 & 2 & 0.64219 & 0.49040 & 0.63363 & 0.49601 & 1.35 \\
4 & 0 & 0.87621 & 0.09542 & 0.86742 & 0.09639 & 1.01 \\
4 & 1 & 0.86511 & 0.28788 & 0.85581 & 0.29088 & 1.09 \\
4 & 2 & 0.84393 & 0.48503 & 0.83369 & 0.49032 & 1.23 \\

\hline
\multicolumn{7}{c}{Electromagnetic perturbations ($s=1$)} \\
\hline
2 & 0 & 0.46285 & 0.09411 & 0.45760 & 0.09500 & 1.15 \\
2 & 1 & 0.44270 & 0.28773 & 0.43654 & 0.29071 & 1.41 \\
2 & 2 & 0.40872 & 0.49571 & 0.40119 & 0.50158 & 1.88 \\
3 & 0 & 0.66402 & 0.09469 & 0.65690 & 0.09562 & 1.08 \\
3 & 1 & 0.64952 & 0.28680 & 0.64174 & 0.28973 & 1.21 \\
3 & 2 & 0.62278 & 0.48669 & 0.61383 & 0.49207 & 1.46 \\
4 & 0 & 0.86213 & 0.09492 & 0.85310 & 0.09586 & 1.06 \\
4 & 1 & 0.85081 & 0.28641 & 0.84127 & 0.28931 & 1.13 \\
4 & 2 & 0.82924 & 0.48269 & 0.81873 & 0.48784 & 1.28 \\

\hline
\multicolumn{7}{c}{Gravitational perturbations ($s=2$)} \\
\hline
2 & 0 & 0.37783 & 0.08819 & 0.37361 & 0.08898 & 1.13 \\
2 & 1 & 0.35205 & 0.27126 & 0.34658 & 0.27394 & 1.58 \\
2 & 2 & 0.30566 & 0.47319 & 0.29850 & 0.47762 & 2.40 \\
3 & 0 & 0.60592 & 0.09186 & 0.59944 & 0.09270 & 1.08 \\
3 & 1 & 0.58983 & 0.27862 & 0.58264 & 0.28130 & 1.23 \\
3 & 2 & 0.56016 & 0.47410 & 0.55169 & 0.47909 & 1.54 \\
4 & 0 & 0.81773 & 0.09328 & 0.80918 & 0.09416 & 1.06 \\
4 & 1 & 0.80573 & 0.28159 & 0.79663 & 0.28433 & 1.14 \\
4 & 2 & 0.78281 & 0.47502 & 0.77271 & 0.47991 & 1.31 \\

\end{tabular}
\end{ruledtabular}
\end{table*}

\begin{figure}[htbp]
  \centering
  \includegraphics[width=0.8\textwidth]{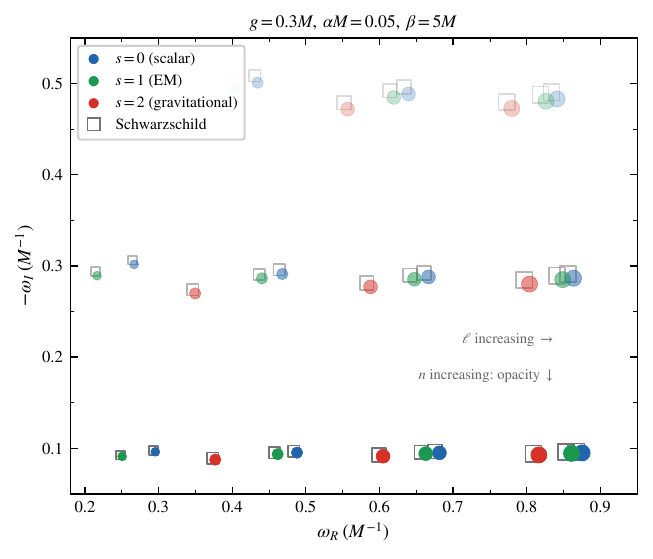}
  \caption{\textcolor{black}{Complex frequency plane $(\omega_R, -\omega_I)$ for the MHDM
           black hole with $g=0.3M$, $\alpha=0.05M$, $\beta=5M$.
           Filled circles: scalar (blue), EM (green), gravitational (red)
           perturbations for $\ell=1$-$4$, $n=0$-$2$.
           Open squares: Schwarzschild reference.
           Point size scales with $\ell$; opacity decreases with $n$.}}
  \label{fig:complex_plane}
\end{figure}

\textcolor{black}{Figure~\ref{fig:complex_plane} summarizes the spectral displacement in the complex-frequency plane.  The MHDM modes are shifted relative to the Schwarzschild reference in a coherent direction across spins, multipoles, and overtones.  This pattern is more informative than a single frequency shift because $g$ and $\alpha$ do not move the modes along exactly the same trajectory in the $(\omega_R,-\omega_I)$ plane.  Thus, although individual observables may be degenerate with Schwarzschild for tuned parameter values, the combined complex spectrum can in principle separate intrinsic magnetic effects from environmental halo effects.}
 
\begin{figure}[htbp]
  \centering
  \includegraphics[width=0.8\textwidth]{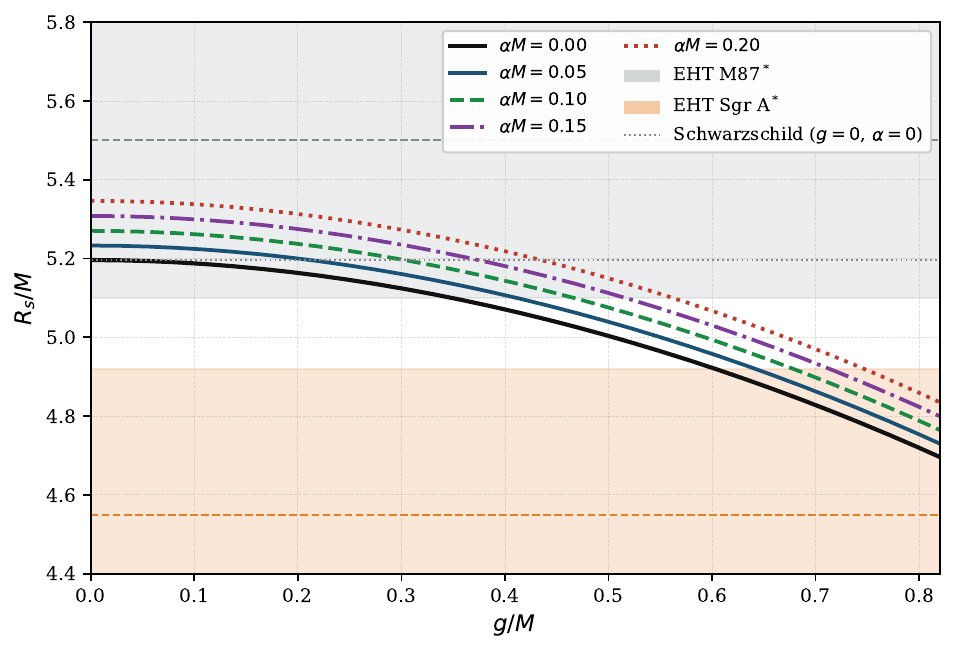}
  \caption{\label{tab:shadow} Shadow radius $R_s/M$ as a function of $g/M$ for
           several values of $\alpha M$, with $\beta=5M$.
           Horizontal bands show the EHT constraints for M87$^*$
           (gray) and Sgr~A$^*$ (orange).}
  \label{fig:shadow}
\end{figure}

In the eikonal regime ($\ell \gg 1$), the QNM spectrum is tied to the unstable photon orbit through \cite{Cardoso:2008bp,Cvetic2016bxi}
\begin{equation}
  \omega_R\simeq \frac{\ell}{R_{sh}},
  \qquad
  R_{sh}=\frac{r_c}{\sqrt{f(r_c)}},
  \label{eq:shadow_qnm}
\end{equation}
where $r_c$ satisfies $2f(r_c)=r_c f'(r_c)$.  Table~\ref{tab:shadow} shows that the magnetic charge reduces the shadow radius, while the halo tends to increase it when the comparison is made at fixed bare mass.  This behavior is consistent with the lapse: the NED term raises $f(r)$ in the strong-field region, whereas the halo lowers it and increases the effective mass scale.  The EHT-inferred ranges for M87$^*$ and Sgr\,A$^*$ therefore provide complementary, although currently broad, constraints.  In particular, shadow information alone cannot remove all parameter degeneracies, but when combined with ringdown frequencies it constrains the NED and halo sectors in different directions.

Table~\ref{tab:shadow} lists $R_s/\mathcal{M}$ as a function of $g$ and 
$\alpha$. The magnetic charge reduces $R_s$ monotonically (from $5.196$ at 
$g=0$ to $4.720$ at $g=0.8\mathcal{M}$ for $\alpha=0$), while the halo raises 
it (from $5.196$ at $\alpha=0$ to $5.347$ at $\alpha/\mathcal{M}=0.20$ for 
$g=0$). Comparing with the EHT constraints for M87$^*$ 
($R_s/\mathcal{M}=5.5\pm0.4$) and Sgr~A$^*$ 
($R_s/\mathcal{M}=4.55^{+0.37}_{-0.17}$), the Sgr~A$^*$ bound favors 
moderate-to-large magnetic charge ($g\gtrsim 0.5\mathcal{M}$ at low $\alpha$), 
while the M87$^*$ bound does not discriminate within the parameter space 
explored.

\begin{table*}[t]
\caption{
Shadow radius $R_s/M$ as a function of $g/M$ and $\alpha M$
for $\beta=5M$. The Schwarzschild value is $3\sqrt{3}\approx5.196$. For comparison, the EHT-inferred characteristic
shadow sizes are \(R_s/M=5.5\pm0.4\) for M87$^*$ and
\(R_s/M=4.55^{+0.37}_{-0.17}\) for Sgr A$^*$.
}
\centering
\small
\setlength{\tabcolsep}{4pt}
\renewcommand{\arraystretch}{1.1}
\begin{ruledtabular}
\begin{tabular}{
S[table-format=1.1]
S[table-format=1.3]
S[table-format=1.3]
S[table-format=1.3]
S[table-format=1.3]
S[table-format=1.3]
}
{$g/M$} & {$\alpha M=0.00$} & {$\alpha M=0.05$} & {$\alpha M=0.10$} &
{$\alpha M=0.15$} & {$\alpha M=0.20$} \\
\colrule
0.0 & 5.196 & 5.233 & 5.270 & 5.308 & 5.347 \\
0.2 & 5.164 & 5.200 & 5.237 & 5.275 & 5.313 \\
0.4 & 5.071 & 5.107 & 5.144 & 5.181 & 5.219 \\
0.6 & 4.923 & 4.958 & 4.994 & 5.030 & 5.067 \\
0.8 & 4.720 & 4.754 & 4.788 & 4.823 & 4.859 \\
\end{tabular}
\end{ruledtabular}
\end{table*}

\section{Energy deposition by neutrino-antineutrino annihilation}
\label{sec:neutrino}

We finally turn to the high-energy process $\nu\bar\nu\rightarrow e^-e^+$.  This channel converts energy carried by neutrinos into an electron-positron plasma and is often invoked in compact-object environments where strong gravity, high temperatures, and relativistic outflows coexist.  In curved spacetime the deposition rate is modified by three geometric effects: gravitational redshift changes the local temperature, null-ray bending changes the angular collision factor, and the spatial metric changes the proper-volume element.  The process is therefore a useful non-wave diagnostic of the same geometry probed by QNMs, shadows, and weak lensing.  Early studies showed that general relativistic effects can substantially amplify the deposition rate near neutron stars and Kerr black holes, emphasizing the importance of strong-field gravity in this process \cite{Salmonson:1999es,Asano:2000ib,Asano:2000dq}.  This line of research has since been generalized to modified-gravity and nonstandard black-hole scenarios, including alternative gravity effects near neutron stars, quintessence-surrounded black holes, Lorentz-violating backgrounds, noncommutative geometries, and Kalb-Ramond gravity, where measurable changes in the annihilation efficiency may have signatures of the underlying gravitational theory \cite{Lambiase:2020iul,Lambiase2021,Khodadi2023,AraujoFilho:2024mvz,Pantig:2025eda,Shi:2023kid,Mannobova:2026jna,Kholmuminov:2026vpv,Alloqulov:2024sns}.  The energy-extraction mechanisms, such as magnetic reconnection, have also been investigated in Lorentz-breaking black-hole spacetimes, which gives a clue about the relevance of strong-gravity halo for relativistic energy-release processes \cite{Carleo:2022qlv}. For the present problem, the advantage of a static and spherically symmetric geometry is that the entire calculation can be reduced to a one-dimensional integral once the lapse function is known.

The NED-Hernquist spacetime is especially interesting for this problem because the two deformations have opposite effects on the lapse.  A magnetic charge raises $f(r)$ outside the horizon and weakens the redshift enhancement of the annihilation kernel.  The Hernquist halo lowers $f(r)$ and therefore strengthens the redshift contribution.  We organize the calculation in two steps.  First, we derive a master formula valid for any static metric of the form \eqref{eq:metric}.  Second, we insert the specific lapse \eqref{eq:lapse} and study the dependence on $g$, $\alpha$, and $\beta$.

\subsection{General neutrino-annihilation formalism}

The starting point is the local energy-deposition rate per unit coordinate time and per unit proper volume.
For a single neutrino flavor pair it is convenient to write it as
\begin{equation}
  \frac{\dd E(r)}{\dd t\,\dd V}
  =2K G_F^2 F(r)
  \iint n(\varepsilon_\nu)\,n(\varepsilon_{\bar\nu})\,(\varepsilon_\nu+\varepsilon_{\bar\nu})\,
  \varepsilon_\nu^3\varepsilon_{\bar\nu}^3\,\dd\varepsilon_\nu\,\dd\varepsilon_{\bar\nu}.
  \label{eq:local_rate_start}
\end{equation}
This expression already exhibits the natural separation of the problem.
The coefficient $K$ contains the weak-interaction information, the thermal distributions $n(\varepsilon)$ encode the neutrino bath, and the function $F(r)$ is a purely geometrical angular factor.
In other words, once $F(r)$ is known from the null geodesics, all remaining geometry enters only through the local temperature redshift.

The weak factor is
\begin{equation}
  K=\frac{1}{6\pi}\left(1\pm 4\sin^2\theta_W+8\sin^4\theta_W\right),
  \label{eq:Kfactor}
\end{equation}
where the upper sign corresponds to electron neutrinos and the lower sign to muon or tau neutrinos.
This coefficient affects only the overall normalization, so it will not alter the relative ordering of the dimensionless curves shown later.

The phase-space distribution is taken to be Fermi-Dirac,
\begin{equation}
  n(\varepsilon)=\frac{2}{h^3}\frac{1}{\exp[\varepsilon/(kT)]+1}.
  \label{eq:FD_distribution}
\end{equation}
The factor of two accounts for spin degeneracy.
Because the integral in Eq.~\eqref{eq:local_rate_start} contains a seventh-order polynomial in the neutrino energies, the final rate acquires the characteristic ninth-power dependence on the local temperature.

Using the standard Fermi integral identity, one obtains the closed result
\begin{equation}
  \frac{\dd E}{\dd t\,\dd V}
  =\frac{21\zeta(5)\pi^4}{h^6}K G_F^2 F(r)(kT)^9.
  \label{eq:local_rate_T9}
\end{equation}
This equation is physically important because it makes the two drivers of the local annihilation rate manifest:
the angular factor $F(r)$ and the steep thermal weight $T^9$.
The latter is the reason why even modest redshift effects can have a large impact on the final deposition efficiency.

To determine $F(r)$ we analyze null rays in the equatorial plane.
The conserved quantities are
\begin{equation}
  E=f(r)\,\dot t,
  \qquad
  L=r^2\dot\phi,
  \qquad
  \dot r^2=E^2-f(r)\frac{L^2}{r^2},
  \label{eq:null_constants}
\end{equation}
where the overdot denotes differentiation with respect to an affine parameter.
These relations show that the entire trajectory information is encoded in the impact parameter $b=L/E$.
In the local orthonormal frame of a static observer at radius $r$, the angle $\theta_r$ between the ray and the tangent direction to the circular orbit satisfies
\begin{equation}
  \cos\theta_r=\frac{b\sqrt{f(r)}}{r}.
  \label{eq:costheta_local}
\end{equation}
This relation has a simple interpretation:
at fixed $b$, a stronger gravitational redshift lowers the locally measured tangential projection and changes the opening of the annihilation cone.

From Eq.~\eqref{eq:costheta_local} one immediately gets
\begin{equation}
  \sin^2\theta_r=1-\frac{b^2f(r)}{r^2}.
  \label{eq:sintheta_local}
\end{equation}
At the neutrinosphere radius $R$ the relevant rays are emitted tangentially, so $\theta_R=0$ and hence
\begin{equation}
  b^2=\frac{R^2}{f(R)}.
  \label{eq:impact_parameter_surface}
\end{equation}
Substituting Eq.~\eqref{eq:impact_parameter_surface} into Eq.~\eqref{eq:sintheta_local} yields the key geometrical relation
\begin{equation}
  x^2\equiv\sin^2\theta_r=1-\frac{R^2}{r^2}\frac{f(r)}{f(R)}.
  \label{eq:x2_general}
\end{equation}
This relation provides the geometric input that determines the local opening angle of the annihilation cone.
It maps the spacetime geometry directly into the local annihilation cone, and it is precisely this map that later transmits the effects of the magnetic charge and dark matter halo into the deposition rate.

The angular factor is defined by averaging the annihilation kernel over the two incident solid angles,
\begin{equation}
  F(r)=\iint\left(1-\bm\Omega_\nu\cdot\bm\Omega_{\bar\nu}\right)^2\dd\Omega_\nu\,\dd\Omega_{\bar\nu}.
  \label{eq:F_definition}
\end{equation}
After carrying out the angular integration one obtains
\begin{equation}
  F(r)=\frac{2\pi^2}{3}(1-x)^4\left(x^2+4x+5\right).
  \label{eq:F_closed}
\end{equation}
Thus the angular dependence of the annihilation kernel is fully determined by the single geometric variable $x$.
The factor becomes large when the geometry allows wide relative angles between neutrinos and antineutrinos, and it decreases when the rays are too strongly collimated.

\subsubsection{Redshifted temperature and integrated rate}

The local temperature is not independent of position.
For a static spacetime the Tolman law gives
\begin{equation}
  T(r)\sqrt{f(r)}=T(R)\sqrt{f(R)}.
  \label{eq:Tolman}
\end{equation}
This shows that the deeper the emission region lies in the gravitational potential, the hotter the local neutrino bath appears to a static observer.
Since the annihilation rate scales as $T^9$, the redshift structure of the metric has a highly amplified effect.

It is convenient to eliminate the local surface temperature in favor of the luminosity at infinity.
Writing the neutrino luminosity at the neutrinosphere as
\begin{equation}
  L(R)=\frac{7\pi ac}{4}R^2T^4(R),
  \label{eq:surface_luminosity}
\end{equation}
and relating it to the asymptotic luminosity through
\begin{equation}
  L_\infty=f(R)L(R),
  \label{eq:luminosity_infinity}
\end{equation}
one finds
\begin{equation}
  T^9(r)=\left(\frac{7\pi ac}{4}\right)^{-9/4}L_\infty^{9/4}R^{-9/2}
  \frac{[f(R)]^{9/4}}{[f(r)]^{9/2}}.
  \label{eq:T9_redshifted}
\end{equation}
The numerator contains the compactness of the neutrinosphere, while the denominator contains the local redshift amplification.
This is exactly the structure that will later explain why lowering the lapse enhances the annihilation rate.

Substituting Eq.~\eqref{eq:T9_redshifted} into Eq.~\eqref{eq:local_rate_T9} gives the local deposition rate in the redshifted form
\begin{equation}
\begin{split}
\frac{\dd E(r)}{\dd t\,\dd V}
&=
\frac{21\zeta(5)\pi^4}{h^6}\,K G_F^2 k^9
\left(\frac{7\pi ac}{4}\right)^{-9/4}
\\
&\quad \times
L_\infty^{9/4} R^{-9/2}
\frac{[f(R)]^{9/4}}{[f(r)]^{9/2}}\,F(r).
\end{split}
\label{eq:local_rate_redshifted}
\end{equation}
This is the master local expression from which both the integrated power and the differential shell profile follow.
It makes the physical content transparent:
all spacetime effects enter through the cone factor $F(r)$ and through the strong inverse power of the lapse.

The proper spatial volume element on a static slice is
\begin{equation}
  \dd V_{\mathrm{prop}}=4\pi r^2\sqrt{g_{rr}}\,\dd r
  =4\pi r^2\frac{\dd r}{\sqrt{f(r)}}.
  \label{eq:proper_volume}
\end{equation}
Integrating the local rate over the exterior region then leads to
\begin{equation}
  \dot Q=\mathcal{C}[f(R)]^{9/4}R^{-9/2}\int_R^{\infty}r^2\frac{F(r)}{[f(r)]^5}\,\dd r,
  \label{eq:Qdot_dimful}
\end{equation}
where $\mathcal{C}$ is a positive constant independent of the geometry, it comes from combining $f^{-9/2}$ in the redshifted temperature with $f^{-1/2}$ from the proper-volume element.
This is the most important structural feature of the global deposition integral.

Introducing the dimensionless radius
\begin{equation}
  y\equiv\frac{r}{R},
  \label{eq:y_def}
\end{equation}
we obtain the normalized integrated rate
\begin{equation}
  \frac{\dot Q}{\dot Q_{\mathrm{Newt}}}
  =3[f(R)]^{9/4}
  \int_1^{\infty}(1-x)^4(x^2+4x+5)\frac{y^2}{[f(Ry)]^5}\,\dd y,
  \label{eq:qratio_master}
\end{equation}
with
\begin{equation}
  x^2=1-\frac{1}{y^2}\frac{f(Ry)}{f(R)}.
  \label{eq:x2_y}
\end{equation}
Equation~\eqref{eq:qratio_master} is the main observable formula used in the first set of figures.
It measures how much the curved spacetime enhances or suppresses the annihilation efficiency relative to the Newtonian limit.

For the local radial distribution it is useful to remove the overall normalization and define the reduced shell profile
\begin{equation}
\begin{split}
\Dred(y)
&\equiv
\frac{R^{5/2}}{\mathcal{C}}\,
\frac{\dd \dot Q}{\dd r}
\\
&=
[f(R)]^{9/4}
(1-x)^4(x^2+4x+5)
\frac{y^2}{[f(Ry)]^{11/4}}.
\end{split}
\label{eq:Dred_general}
\end{equation}
This quantity is proportional to the true $\dd\dot Q/\dd r$ and preserves all relative ordering information.
It is therefore the natural object to plot when one wants to compare how different parameters redistribute the deposited energy in radius.

To make the numerical analysis transparent, we introduce the dimensionless variables
\begin{equation}
\begin{split}
\rho &\equiv \frac{r}{M}, \qquad
\rho_R \equiv \frac{R}{M}, \\
\bar g &\equiv \frac{g}{M}, \qquad
\bar\alpha \equiv \frac{\alpha}{M}, \qquad
\bar\beta \equiv \frac{\beta}{M}.
\end{split}
\label{eq:dimensionless_defs}
\end{equation}
In terms of these variables the lapse becomes
\begin{equation}
\begin{split}
\fM(\rho)
&=
1-\frac{2}{\rho}
+\frac{\bar g^2}{\rho(\rho+\bar g)}
-\frac{2\bar\alpha\rho}{(\rho+\bar\beta)^2}.
\end{split}
\label{eq:lapse_dimensionless}
\end{equation}

This form is much better suited for parameter scans because it isolates the physically relevant control parameters.
The sign structure is already revealing:
the magnetic term is positive outside the horizon and tends to raise the lapse, while the Hernquist term is negative and tends to lower it.

In the MHDM spacetime, Eqs.~\eqref{eq:qratio_master} and \eqref{eq:Dred_general} become
\begin{equation}
  \frac{\dot Q}{\dot Q_{\mathrm{Newt}}}
  =3[\fM(\rho_R)]^{9/4}
  \int_1^{\infty}(1-x)^4(x^2+4x+5)\frac{y^2}{[\fM(\rho_R y)]^5}\,\dd y,
  \label{eq:qratio_MHDM}
\end{equation}
with
\begin{equation}
  x^2=1-\frac{1}{y^2}\frac{\fM(\rho_R y)}{\fM(\rho_R)},
  \label{eq:x2_MHDM}
\end{equation}
and
\begin{equation}
  \Dred(y)=[\fM(\rho_R)]^{9/4}(1-x)^4(x^2+4x+5)\frac{y^2}{[\fM(\rho_R y)]^{11/4}}.
  \label{eq:Dred_MHDM}
\end{equation}
These are the final working equations used in the numerical plots.
They are structurally identical to the general master formulas, but now the role of the magnetic charge and halo strength is explicit.


Because the spacetime contains two intrinsic deformation amplitudes and one halo
scale, a single one-parameter plot cannot capture the relevant physics.  We
therefore separate the analysis into two complementary sectors.  In the first
sector, the Hernquist parameters are kept fixed and the magnetic charge is
varied.  This isolates the effect of the nonlinear-electrodynamics magnetic
monopole.  In the second sector, the magnetic charge is fixed and the Hernquist
amplitude is varied, isolating the effect of the surrounding dark-matter halo.
Throughout this section we use
\begin{equation}
  \bar\beta=2 ,
  \label{eq:beta_choice_neutrino}
\end{equation}
and define the two scans by
\begin{equation}
\begin{aligned}
\text{charge sector:}\qquad
& \bar g=0,\;0.4,\;0.8,\;1.2,
\qquad
\bar\alpha=0.10,
\\
\text{halo sector:}\qquad
& \bar\alpha=0,\;0.05,\;0.10,\;0.15,
\qquad
\bar g=0.50 .
\end{aligned}
\label{eq:scan_sectors_neutrino}
\end{equation}
This split is physically useful because the two deformations enter the lapse
with opposite signs.  The magnetic contribution raises \(f(r)\) outside the
horizon, while the Hernquist contribution lowers it.  Since the total
deposition integral contains the strong weight \(f^{-5}\), this sign difference
already suggests that the magnetic charge should suppress the annihilation
efficiency, whereas the halo should enhance it.

For the integrated deposition rate we use \(4\leq R/M\leq 7\), so that the
neutrinosphere lies safely outside the horizon for all displayed parameter
choices.  For the local shell profile we choose \(R/M=3.2\).  This still probes
the strong-field region but avoids parameter points for which the tangential-ray
construction would place part of the near-surface cone too close to the
photon-sphere threshold.

\begin{table*}[t]
\caption{\label{tab:params_neutrino}
Parameter sets used in the neutrino-annihilation analysis.  The quoted outer
horizon radii confirm that the adopted neutrinosphere radii lie in the exterior
region.}
\begin{ruledtabular}
\begin{tabular}{cccc}
Sector & Varied parameter & Fixed values & \(r_h/M\) \\
\hline
charge & \(\bar g=0.0\) & \((\bar\alpha,\bar\beta)=(0.10,2)\) & 2.051 \\
charge & \(\bar g=0.4\) & \((\bar\alpha,\bar\beta)=(0.10,2)\) & 1.982 \\
charge & \(\bar g=0.8\) & \((\bar\alpha,\bar\beta)=(0.10,2)\) & 1.799 \\
charge & \(\bar g=1.2\) & \((\bar\alpha,\bar\beta)=(0.10,2)\) & 1.504 \\
\hline
halo & \(\bar\alpha=0.00\) & \((\bar g,\bar\beta)=(0.50,2)\) & 1.896 \\
halo & \(\bar\alpha=0.05\) & \((\bar g,\bar\beta)=(0.50,2)\) & 1.921 \\
halo & \(\bar\alpha=0.10\) & \((\bar g,\bar\beta)=(0.50,2)\) & 1.947 \\
halo & \(\bar\alpha=0.15\) & \((\bar g,\bar\beta)=(0.50,2)\) & 1.973 \\
\end{tabular}
\end{ruledtabular}
\end{table*}

The horizon trend in Table~\ref{tab:params_neutrino} already reflects the
competing roles of the two sectors.  Increasing the magnetic charge decreases
the horizon radius, whereas increasing the Hernquist amplitude moves the outer
horizon slightly outward.  The same competition controls the energy deposition
rate, as shown below.

\subsubsection{Charge-sector suppression}

\begin{figure}[t]
  \centering
  \includegraphics[width=0.80\textwidth]{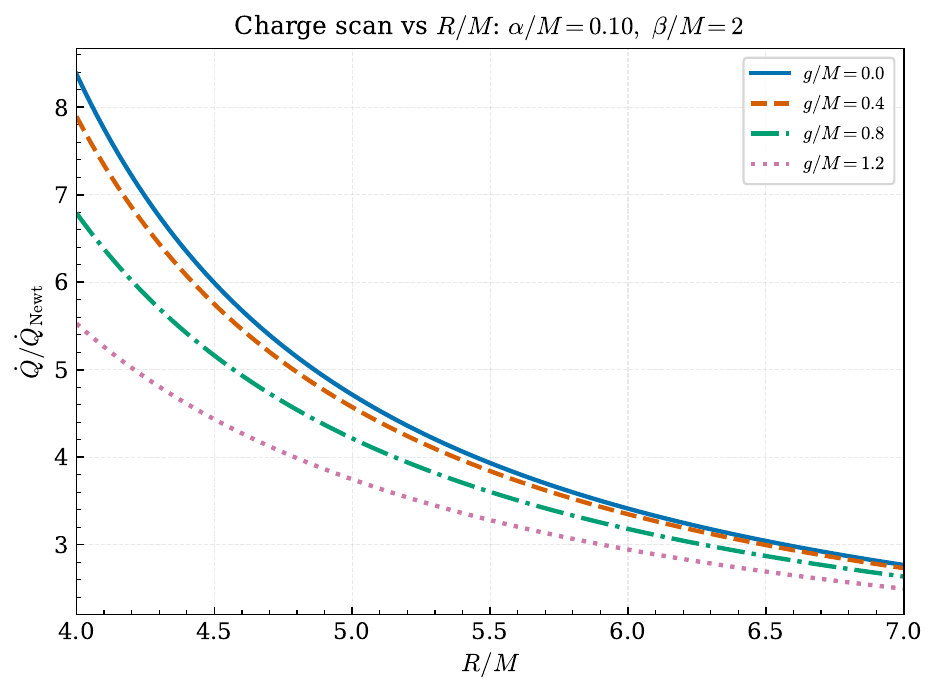}
  \caption{Charge-sector dependence of the normalized energy-deposition rate
  \(\dot Q/\dot Q_{\rm Newt}\) on the neutrinosphere radius \(R/M\), at fixed
  \((\bar\alpha,\bar\beta)=(0.10,2)\).  Increasing the magnetic charge lowers
  the deposition efficiency.  The suppression is strongest for smaller \(R/M\),
  where the annihilation region samples the deepest part of the gravitational
  potential.}
  \label{fig:qratio_R_charge}
\end{figure}

We first vary the magnetic charge at fixed \((\bar\alpha,\bar\beta)=(0.10,2)\).
Figure~\ref{fig:qratio_R_charge} shows the normalized deposition rate as a
function of \(R/M\).

The ordering of the curves in Fig.~\ref{fig:qratio_R_charge} is monotonic.  The
reason is direct: the magnetic term in the lapse,
\[
  \frac{\bar g^2}{\rho(\rho+\bar g)},
\]
is positive outside the horizon and therefore raises \(f(r)\).  Since the
integrated rate contains \(f^{-5}\), a larger lapse reduces the redshift
amplification of the annihilation kernel.  The effect is cumulative over the
radial integration domain and therefore remains visible even at moderately
large \(R/M\).

\begin{figure}[t]
  \centering
  \includegraphics[width=0.80\textwidth]{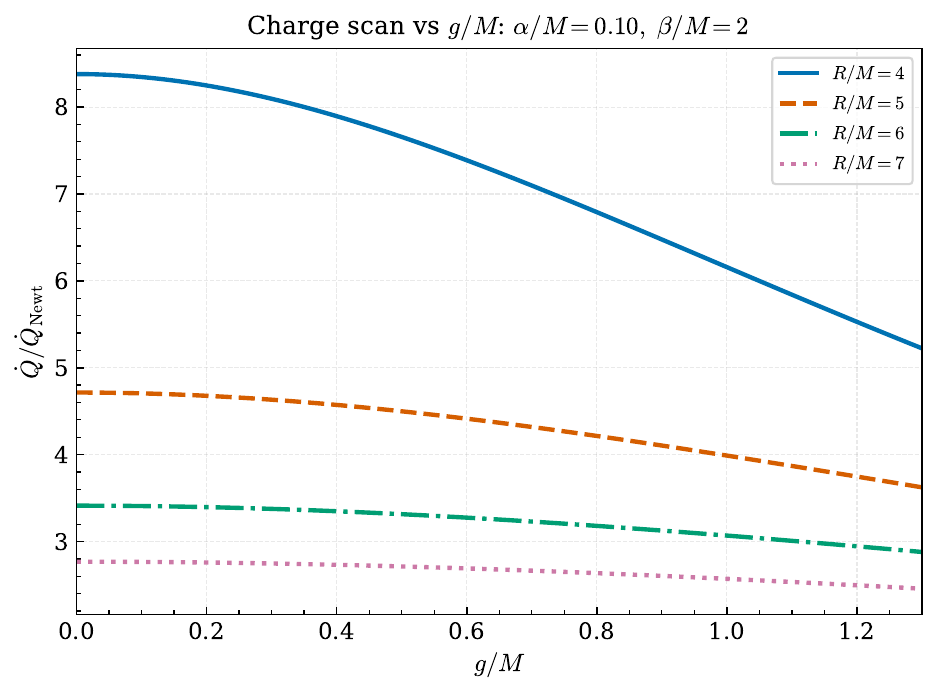}
  \caption{Charge-sector dependence of \(\dot Q/\dot Q_{\rm Newt}\) on the
  magnetic charge \(\bar g=g/M\), at fixed \((\bar\alpha,\bar\beta)=(0.10,2)\).
  Each curve corresponds to a fixed neutrinosphere radius.  The monotonic
  decrease with \(\bar g\) confirms that the NED magnetic charge suppresses the
  relativistic enhancement of the neutrino-annihilation channel.}
  \label{fig:qratio_g_charge}
\end{figure}

The same result is displayed from a complementary viewpoint in
Fig.~\ref{fig:qratio_g_charge}, where \(R/M\) is fixed and the charge is varied.

At \(R/M=4\), the normalized rate decreases from
\[
  \frac{\dot Q}{\dot Q_{\rm Newt}}\simeq 8.3801
  \quad \text{at} \quad \bar g=0
\]
to
\[
  \frac{\dot Q}{\dot Q_{\rm Newt}}\simeq 5.5281
  \quad \text{at} \quad \bar g=1.2 .
\]
Thus, for the adopted halo strength, the magnetic sector can reduce the
deposition efficiency by roughly one third in the strong-field regime.

\subsubsection{Halo-sector enhancement}

\begin{figure}[t]
  \centering
  \includegraphics[width=0.80\textwidth]{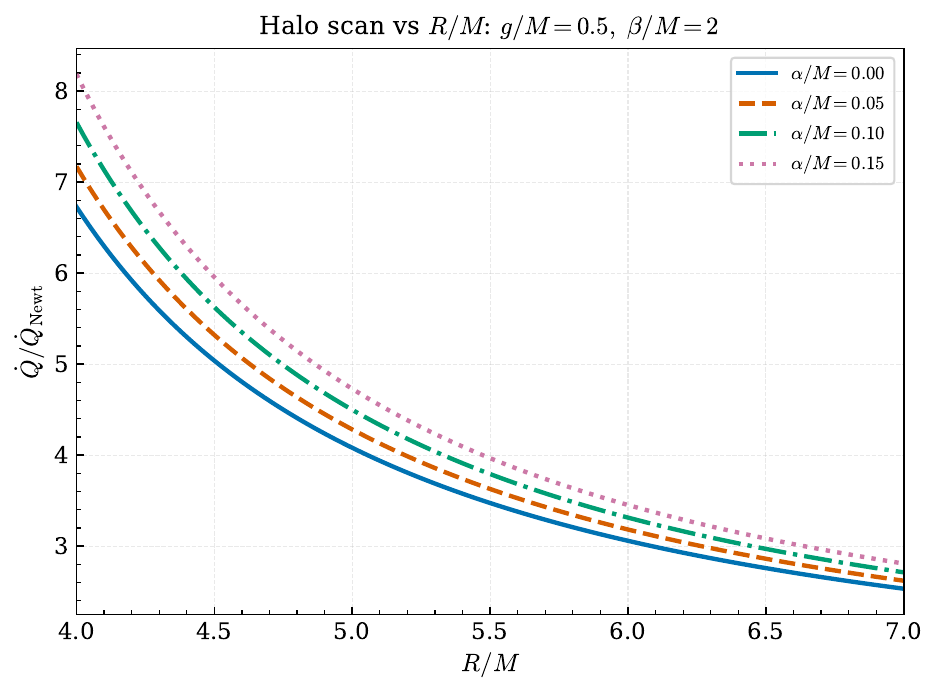}
  \caption{Halo-sector dependence of \(\dot Q/\dot Q_{\rm Newt}\) on \(R/M\),
  at fixed \((\bar g,\bar\beta)=(0.50,2)\).  In contrast to the magnetic
  charge, increasing the Hernquist amplitude enhances the deposition rate.  The
  enhancement is again strongest at smaller \(R/M\), where the redshift and
  ray-bending effects are largest.}
  \label{fig:qratio_R_halo}
\end{figure}

We now fix \(\bar g=0.50\) and vary the Hernquist amplitude \(\bar\alpha\).
Figure~\ref{fig:qratio_R_halo} shows the dependence on \(R/M\).

The trend is opposite to that of the charge sector because the Hernquist
contribution to the lapse is negative,
\[
  -\frac{2\bar\alpha\rho}{(\rho+\bar\beta)^2}.
\]
Increasing \(\bar\alpha\) lowers \(f(r)\), strengthens the Tolman redshift
factor, and enhances the \(f^{-5}\) weight in the global integral.  The halo
therefore acts as an environmental amplifier of the neutrino-annihilation
efficiency.

\begin{figure}[t]
  \centering
  \includegraphics[width=0.80\textwidth]{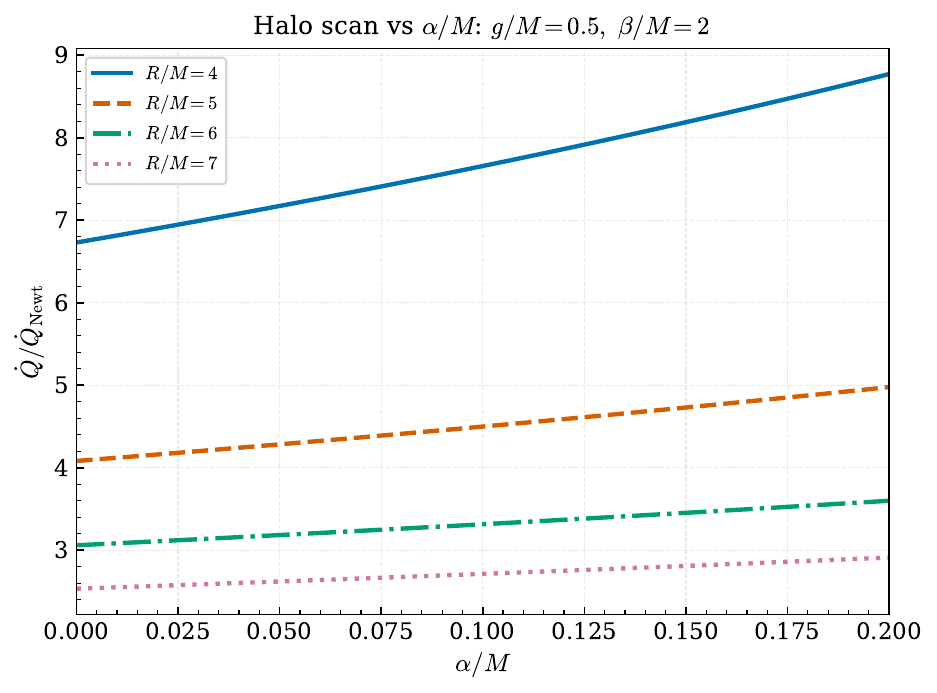}
  \caption{Halo-sector dependence of \(\dot Q/\dot Q_{\rm Newt}\) on the
  Hernquist amplitude \(\bar\alpha=\alpha/M\), at fixed
  \((\bar g,\bar\beta)=(0.50,2)\).  The nearly monotonic increase confirms that
  the dark-matter halo strengthens the relativistic deposition efficiency by
  lowering the lapse in the relevant integration domain.}
  \label{fig:qratio_alpha_halo}
\end{figure}

Figure~\ref{fig:qratio_alpha_halo} makes the same conclusion explicit by
plotting the deposition rate directly as a function of \(\bar\alpha\).

At \(R/M=4\), the rate increases from
\[
  \frac{\dot Q}{\dot Q_{\rm Newt}}\simeq 6.7298
  \quad \text{at} \quad \bar\alpha=0
\]
to
\[
  \frac{\dot Q}{\dot Q_{\rm Newt}}\simeq 8.1880
  \quad \text{at} \quad \bar\alpha=0.15 .
\]
The halo contribution therefore compensates, and can overcompensate, the
magnetic suppression depending on the location of the neutrinosphere.

\subsubsection{Local shell profiles}

The integrated rate determines the total deposited energy, but it does not show
where the energy is deposited.  For this purpose we use the reduced shell
profile \(\mathcal D_{\rm red}(y)\), which is proportional to
\(\mathrm d\dot Q/\mathrm dr\) after removing an overall constant common to all
curves.

\begin{figure}[t]
  \centering
  \includegraphics[width=0.80\textwidth]{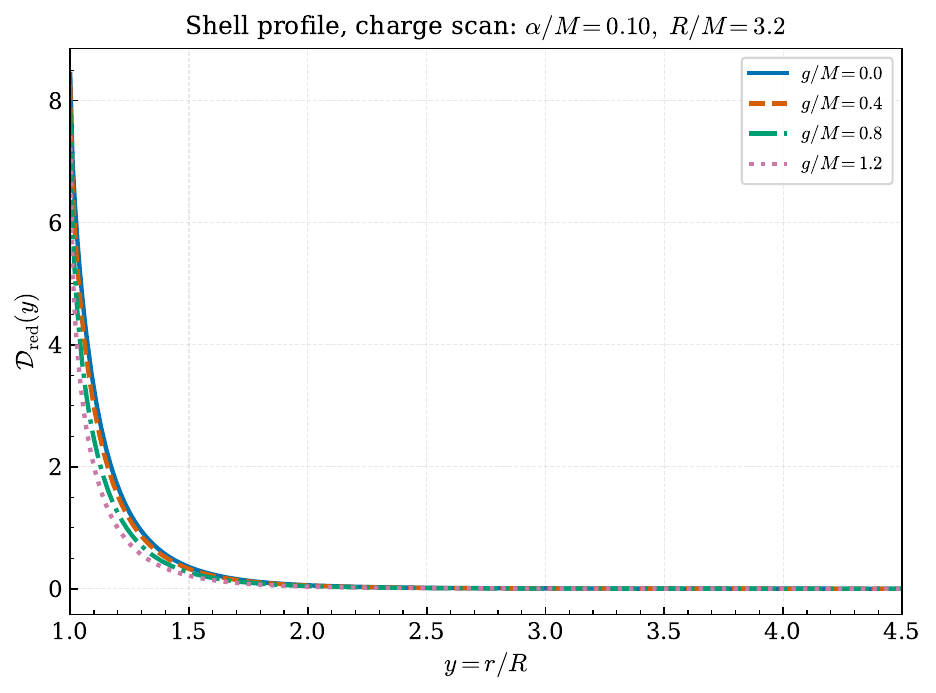}
  \caption{Reduced local shell profile \(\mathcal D_{\rm red}(y)\) in the
  charge sector, at fixed \((\bar\alpha,\bar\beta)=(0.10,2)\) and \(R/M=3.2\).
  The profile is largest close to the neutrinosphere and decreases rapidly
  outward.  Increasing \(\bar g\) suppresses the profile at every radius,
  confirming locally the same trend found for the integrated rate.}
  \label{fig:dqdr_charge}
\end{figure}

The charge-sector shell profile is shown in Fig.~\ref{fig:dqdr_charge}.

\begin{figure}[t]
  \centering
  \includegraphics[width=0.80\textwidth]{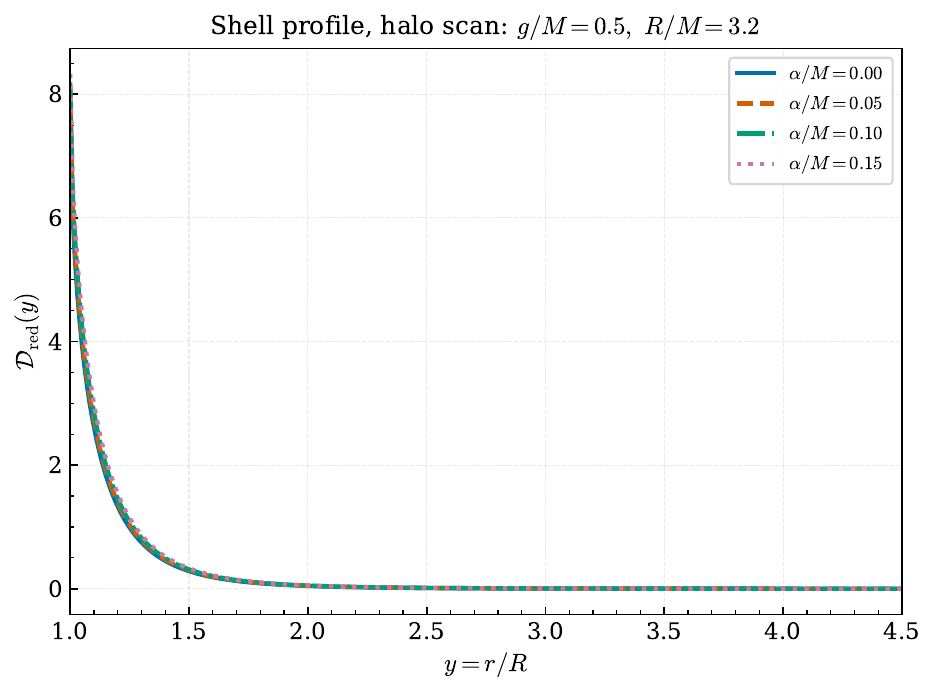}
  \caption{Reduced local shell profile \(\mathcal D_{\rm red}(y)\) in the halo
  sector, at fixed \((\bar g,\bar\beta)=(0.50,2)\) and \(R/M=3.2\).  Increasing
  the Hernquist amplitude enhances the local deposition profile.  The effect is
  strongest near the neutrinosphere, where the redshift and angular-cone
  corrections are most important.}
  \label{fig:dqdr_halo}
\end{figure}

The halo-sector shell profile is shown in Fig.~\ref{fig:dqdr_halo}.

Both Figs.~\ref{fig:dqdr_charge} and \ref{fig:dqdr_halo} show that most of the
deposited energy is generated close to the neutrinosphere.  This is expected
because, at \(y=1\), one has \(x=0\) and therefore
\[
  (1-x)^4(x^2+4x+5)=5 ,
\]
so the angular kernel is large at the emitting surface.  As \(y\) increases,
\(x\) grows and the angular factor decreases, while the redshift enhancement
also weakens.  The radial profile therefore falls rapidly outward.

\subsubsection{Lapse mechanism}

\begin{figure}[t]
  \centering
  \includegraphics[width=0.80\textwidth]{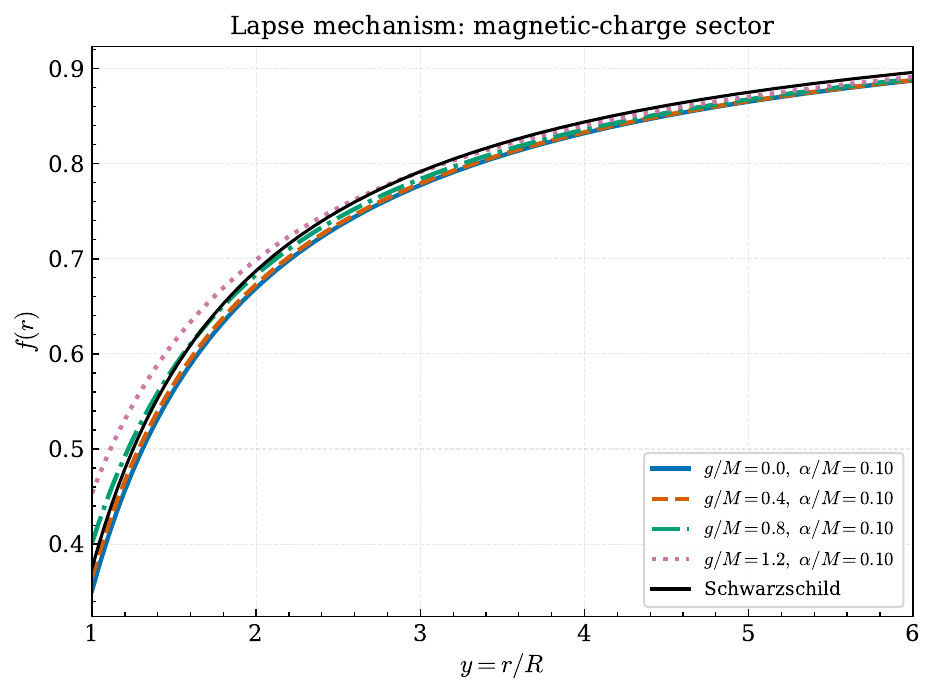}
  \caption{Lapse function \(f(r)\) in the charge sector at fixed
  \((\bar\alpha,\bar\beta)=(0.10,2)\), plotted against \(y=r/R\) with
  \(R/M=3.2\).  Larger magnetic charge raises the lapse outside the horizon.
  This weakens the \(f^{-5}\) weight in the deposition integral and explains
  the suppression observed in Figs.~\ref{fig:qratio_R_charge},
  \ref{fig:qratio_g_charge}, and \ref{fig:dqdr_charge}.}
  \label{fig:lapse_charge}
\end{figure}

\begin{figure}[t]
  \centering
  \includegraphics[width=0.80\textwidth]{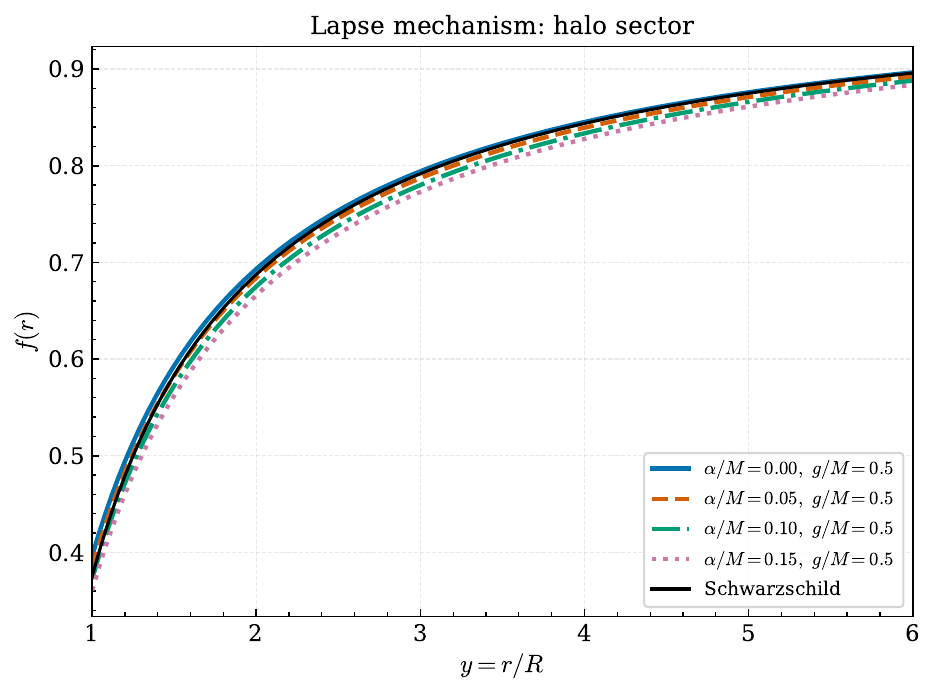}
  \caption{Lapse function \(f(r)\) in the halo sector at fixed
  \((\bar g,\bar\beta)=(0.50,2)\), plotted against \(y=r/R\) with \(R/M=3.2\).
  Larger Hernquist amplitude lowers the lapse in the strong-field region.  This
  enhances the redshift contribution and explains the increase of the deposition
  rate in Figs.~\ref{fig:qratio_R_halo}, \ref{fig:qratio_alpha_halo}, and
  \ref{fig:dqdr_halo}.}
  \label{fig:lapse_halo}
\end{figure}

The opposite behaviour of the charge and halo sectors can be understood most
clearly by plotting the lapse function itself.  Figures~\ref{fig:lapse_charge}
and \ref{fig:lapse_halo} display \(f(r)\) in the two sectors.

These plots show that the sign of the lapse deformation is the organizing
principle of the entire neutrino-annihilation sector.  Any deformation that raises the lapse suppresses the rate; any deformation 
that lowers the lapse enhances it.

\subsubsection{Cumulative deposition fraction}

To quantify where the deposited energy is produced, we define the cumulative
fraction
\begin{equation}
\Phi(y)
\equiv
\frac{\dot Q(<y)}{\dot Q_{\rm total}}
=
\frac{
3[\mathcal f(\rho_R)]^{9/4}
\displaystyle\int_1^y
(1-x)^4(x^2+4x+5)
\frac{y'^{\,2}}{[\mathcal f(\rho_R y')]^5}\,\mathrm dy'
}{
\dot Q/\dot Q_{\rm Newt}
}.
\label{eq:Phi_def}
\end{equation}
This quantity measures the fraction of the total deposition generated between
the neutrinosphere and the radius \(r=yR\).

\begin{figure}[t]
  \centering
  \includegraphics[width=0.60\textwidth]{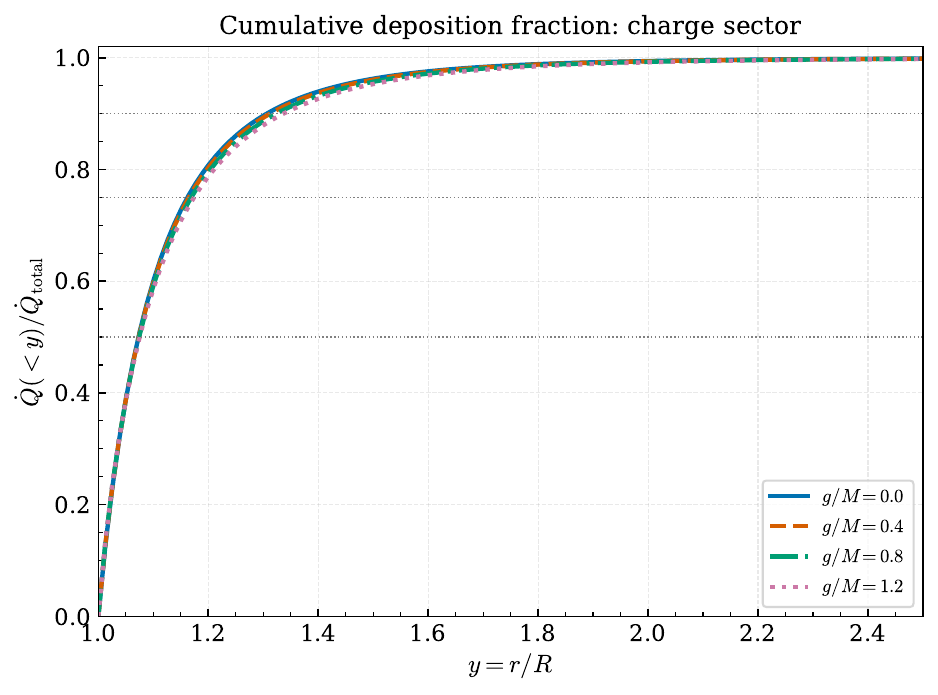}
  \caption{Cumulative deposition fraction \(\Phi(y)\) in the charge sector.
  Horizontal dotted lines mark the \(50\%\), \(75\%\), and \(90\%\) levels.
  The curves are nearly coincident, showing that the magnetic charge changes
  mainly the amplitude of the deposition rate rather than the radial location at
  which the energy is generated.}
  \label{fig:cumulative_charge}
\end{figure}

The charge-sector cumulative fraction is shown in
Fig.~\ref{fig:cumulative_charge}.

\begin{figure}[t]
  \centering
  \includegraphics[width=0.60\textwidth]{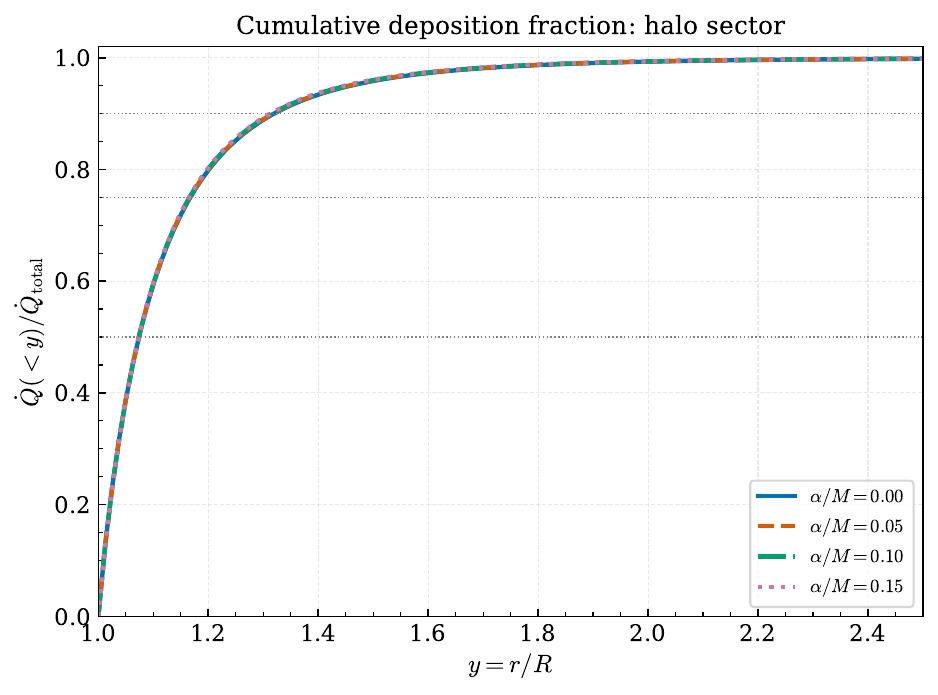}
  \caption{Cumulative deposition fraction \(\Phi(y)\) in the halo sector.
  As in the charge sector, the curves are nearly universal.  The Hernquist halo
  strongly modifies the total amplitude of \(\dot Q\), but it barely shifts the
  radial region where the deposition is produced.}
  \label{fig:cumulative_halo}
\end{figure}

The corresponding halo-sector result is given in
Fig.~\ref{fig:cumulative_halo}.

A useful single diagnostic is the half-deposition radius \(y_{1/2}\), defined by
\[
  \Phi(y_{1/2})=\frac{1}{2}.
\]
For all parameter choices considered here, we find
\[
  y_{1/2}\simeq 1.075 .
\]
Thus approximately half of the total deposited energy is generated within only
about \(7.5\%\) of the neutrinosphere radius.  This near-universality implies
that the NED-Hernquist parameters primarily change the overall amplitude of the
annihilation rate, not the radial location of the energy release.

\subsubsection{Two-parameter map and halo scale}

\begin{figure}[t]
  \centering
  \includegraphics[width=0.60\textwidth]{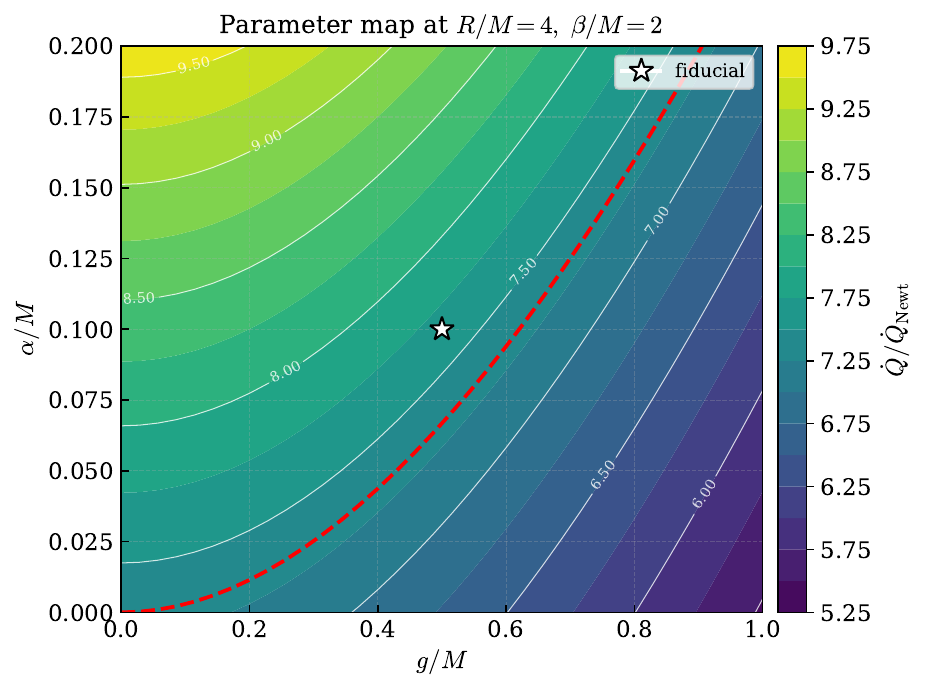}
  \caption{Two-parameter map of \(\dot Q/\dot Q_{\rm Newt}\) in the
  \((g/M,\alpha/M)\) plane at fixed \(R/M=4\) and \(\beta/M=2\).  The filled
  contours show the normalized deposition rate.  The dashed red contour marks
  the Schwarzschild-equivalent value.  The anti-diagonal orientation of the
  contours reflects the competition between the magnetic charge, which
  suppresses the rate, and the Hernquist halo, which enhances it.}
  \label{fig:param_map}
\end{figure}

The preceding scans vary one parameter at a time.  To visualize the full
competition between magnetic charge and halo strength, Fig.~\ref{fig:param_map}
shows the integrated rate in the \((g/M,\alpha/M)\) plane at fixed
\(R/M=4\) and \(\beta/M=2\).

The red dashed curve in Fig.~\ref{fig:param_map} identifies parameter
combinations that reproduce the Schwarzschild deposition rate.  Along this
curve, the suppression induced by the magnetic charge is compensated by the
enhancement induced by the halo.  Therefore, the neutrino-annihilation channel
alone cannot uniquely determine both \(g/M\) and \(\alpha/M\).  It must be
combined with shadow, ringdown, or lensing observables to break this residual
degeneracy.

\begin{figure}[t]
  \centering
  \includegraphics[width=0.70\textwidth]{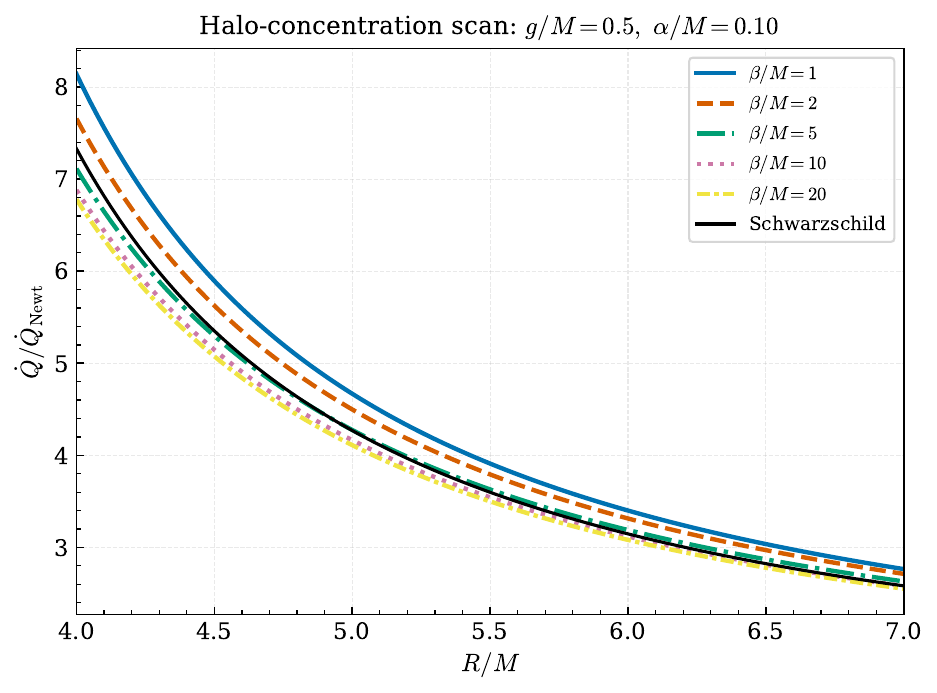}
  \caption{Normalized deposition rate as a function of halo concentration at 
  fixed $(g/\mathcal{M},\alpha/\mathcal{M})=(0.5,0.10)$. Smaller 
  $\beta/\mathcal{M}$ yields a more concentrated halo and a larger enhancement; 
  larger $\beta/\mathcal{M}$ yields a more diffuse halo and the curves tend to 
  the Schwarzschild reference.}
  \label{fig:beta_scan}
\end{figure}

Finally, Fig.~\ref{fig:beta_scan} shows the effect of varying the halo scale 
$\beta/\mathcal{M}$ at fixed $(g/\mathcal{M},\alpha/\mathcal{M})=(0.5,0.10)$.
The $\beta/\mathcal{M}$ scan shows that the neutrino-annihilation channel is 
most sensitive to compact halos. If the halo is concentrated around the central 
object, its negative lapse contribution acts precisely where the integrand is 
largest; if the halo is diffuse, the local strong-field effect weakens and the 
deposition rate approaches the Schwarzschild curve. Any deformation that raises the lapse suppresses the rate; any deformation that 
lowers the lapse enhances it. The cumulative-fraction analysis shows that the 
energy release is always localized near the neutrinosphere, with 
$y_{1/2}\simeq 1.075$. The NED-Hernquist parameters therefore reshape the 
amplitude of $\dot{Q}$ rather than the radial location of the deposition. 
Neutrino-antineutrino annihilation thus provides an independent high-energy 
constraint on the same lapse competition probed by the shadow, ringdown and 
lensing channels.

\begin{table*}[t]
\caption{\label{tab:neutrino_summary}
Normalized neutrino-antineutrino annihilation rate
\(\dot Q/\dot Q_{\rm Newt}\) for representative neutrinosphere radii, with
\(\bar\beta=2\).  The last column gives the half-deposition radius
\(y_{1/2}\), evaluated at \(R/M=4\).}
\begin{ruledtabular}
\begin{tabular}{cccccc}
Sector & Varied parameter & \(R/M=4\) & \(R/M=5\) & \(R/M=6\) & \(y_{1/2}\) \\
\hline
\multicolumn{6}{c}{Charge scan: \(\bar\alpha=0.10,\ \bar\beta=2\)}\\
 & \(\bar g=0.0\) & 8.3801 & 4.7146 & 3.4124 & 1.075 \\
 & \(\bar g=0.4\) & 7.8958 & 4.5722 & 3.3482 & 1.075 \\
 & \(\bar g=0.8\) & 6.7910 & 4.2148 & 3.1798 & 1.075 \\
 & \(\bar g=1.2\) & 5.5281 & 3.7472 & 2.9445 & 1.076 \\
\hline
\multicolumn{6}{c}{Halo scan: \(\bar g=0.50,\ \bar\beta=2\)}\\
 & \(\bar\alpha=0.00\) & 6.7298 & 4.0815 & 3.0595 & 1.075 \\
 & \(\bar\alpha=0.05\) & 7.1727 & 4.2831 & 3.1835 & 1.075 \\
 & \(\bar\alpha=0.10\) & 7.6570 & 4.4987 & 3.3144 & 1.075 \\
 & \(\bar\alpha=0.15\) & 8.1880 & 4.7296 & 3.4529 & 1.075 \\
\end{tabular}
\end{ruledtabular}
\end{table*}

\begin{table}[t]
\caption{\label{tab:neutrino_beta}
Halo-scale dependence of \(\dot Q/\dot Q_{\rm Newt}\) at fixed
\((g/M,\alpha/M)=(0.5,0.10)\) and \(R/M=4\).}
\begin{ruledtabular}
\begin{tabular}{ccccc}
\(\beta/M\) & 1.0 & 2.0 & 5.0 & 10.0 \\
\hline
\(\dot Q/\dot Q_{\rm Newt}\) & 8.1414 & 7.6570 & 7.1133 & 6.8824 \\
\end{tabular}
\end{ruledtabular}
\end{table}
\section{Parameter degeneracy and multi-observable constraints}
\label{sec:degeneracy}

The numerical results presented in Secs.~\ref{sec:shadow}-\ref{sec:perturbations} show that the NED charge $g$ and the halo amplitude $\alpha$ produce competing shifts in individual observables. For any fixed value of $\beta/M$ this, it raises a natural question: given a measured value of one observable, can the underlying parameters $(g/M,\,\alpha/M)$ be uniquely determined, or do degenerate combinations exist?  The answer to this question is by mapping the four observables studied
in this paper; the shadow radius $R_{ sh}$, the fundamental QNM
frequency $M\omega_R$, the neutrino-annihilation rate
$\dot{Q}/\dot{Q}_{ Newt}$, and the asymptotic deflection angle
$\hat{\theta}_\infty$, simultaneously onto the parameter plane.

\begin{figure}[t]
  \centering
  \includegraphics[width=0.8 \textwidth]{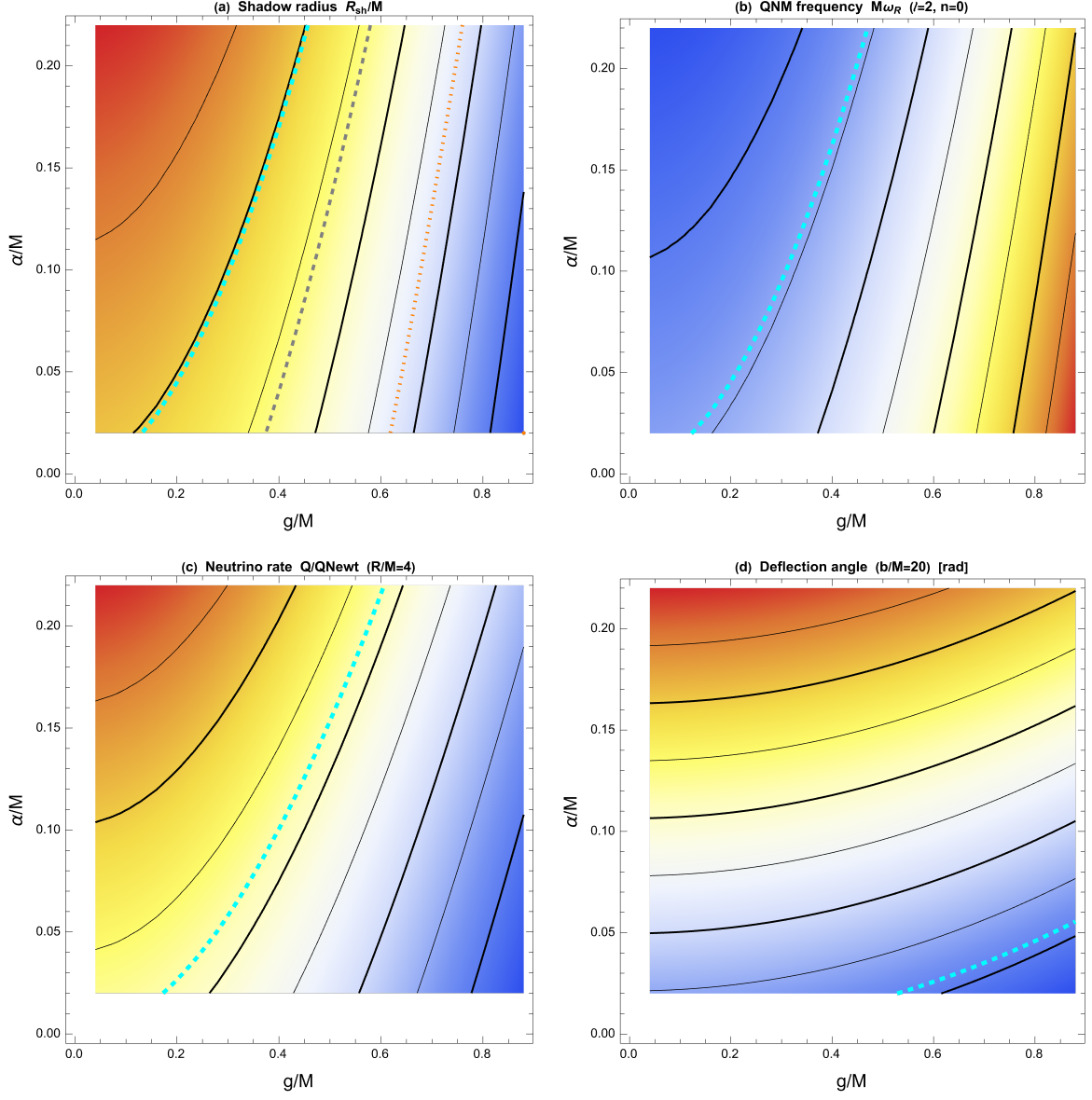}
  \caption{Contour maps of four observables in the $(g/M,\,\alpha/M)$
  parameter plane at fixed $\beta/M = 5$.
  \textbf{(a)}~Shadow radius $R_{ sh}/M$.
  \textbf{(b)}~Fundamental QNM frequency $M\omega_R$ (scalar
  perturbations, $\ell=2$, $n=0$).
  \textbf{(c)}~Normalized neutrino-annihilation rate
  $\dot{Q}/\dot{Q}_{ Newt}$ at neutrinosphere radius $R/M=4$.
  \textbf{(d)}~Asymptotic weak-deflection angle $\hat{\theta}_\infty$
  at impact parameter $b/M = 20$.
  In all panels, black contours show lines of constant observable value
  and the cyan dashed contour marks the Schwarzschild reference.
  In panel~(a), the gray dashed and orange dotted bands indicate the
  EHT constraints for M87$^*$ ($R_{ sh}/M = 5.5\pm0.4$) and
  Sgr~A$^*$ ($R_{ sh}/M = 4.55^{+0.37}_{-0.17}$), respectively.
  The non-parallelism of the contour families across panels
  demonstrates that the four observables carry complementary information
  on $(g/M,\,\alpha/M)$ and that their combination breaks the
  degeneracy that any single observable leaves unresolved.}
  \label{fig:degeneracy}
\end{figure}

Figure~\ref{fig:degeneracy} shows contour plots of each observable
as a function of $(g/M,\,\alpha/M)$ at a fixed $\beta/M = 5$.
Each family of contours defines a one-parameter degeneracy locus
in the parameter plane: along any single contour, the corresponding observable cannot distinguish between different $(g,\alpha)$ combinations.
The key structural result is that the four contour families are
mutually non-parallel. This geometric property implies that no pair of observables shares the same degeneracy direction, so their combination generically breaks the remaining freedom and allows both parameters to be determined simultaneously.

The physical origin of the non-parallelism can be traced to the
radial weighting of each observable.
The shadow radius is governed by the unstable photon orbit at
$r_{ ph} \approx 3\mathcal{M}$, where $\mathcal{M} = M + \alpha$
is the asymptotic mass.
Because $\alpha$ contributes to $\mathcal{M}$ through the $1/r$
sector of the lapse, $R_{ sh}$ is primarily sensitive to $\alpha$
and only weakly to $g$ at fixed $\mathcal{M}$, producing nearly
vertical contours in panel~(a). By contrast, the QNM frequency is controlled by the height of the effective potential barrier, which is raised by the compact NED term $g^2/[r(r+g)]$ and lowered by the halo contribution. Since the NED correction is short-range and peaks near the horizon,
$M\omega_R$ is more sensitive to changes $g$ than to $\alpha$ at fixed $\mathcal{M}$, producing contours in panel~(b) with a slope
approximately orthogonal to those in panel~(a).
This near-orthogonality is precisely the condition that makes the
shadow-QNM combination most efficient at resolving the two-parameter
degeneracy.
The neutrino-annihilation rate and the weak-deflection angle probe
complementary radial regions. The rate $\dot{Q}/\dot{Q}_{ Newt}$ depends on the lapse throughout the exterior region between the neutrinosphere and infinity, weighting both the compact NED sector and the extended halo contribution with intermediate radial leverage.
Its contours in panel~(c) have an intermediate slope that is neither parallel to those of $R_{ sh}$ nor to those of $M\omega_R$, confirming that it carries independent information about both parameters.
The asymptotic deflection angle $\hat{\theta}_\infty$ at $b/M = 20$ is dominated by the Newtonian term $4\mathcal{M}/b$, which depends primarily on $\alpha$ through the asymptotic mass.
Its contours in panel~(d) are therefore nearly horizontal and
structurally similar to those of $R_{ sh}$, indicating that
$\hat{\theta}_\infty$ and $R_{ sh}$ are partially degenerate
with each other. However, both are complementary to $M\omega_R$ and
$\dot{Q}/\dot{Q}_{ Newt}$.
These results demonstrate that a combined analysis is necessary and sufficient to constrain the NED-Hernquist parameter space.
In practice, a measurement of $R_{ sh}$ primarily constrains
$\alpha/M$ at fixed asymptotic mass, while a ringdown observation
of $M\omega_R$ constrains $g/M$ independently.
The neutrino channel adds a third constraint that is sensitive to the product $g\alpha$ through the competition of the two lapse
deformations. Although current observational uncertainties are too large to fully exploit this degeneracy breaking structure,  the EHT fractional
precision on $R_{sh}$ is of order $7\%$ for M87$^*$ and Sgr~A$^*$,
while ringdown measurements from LISA and next-generation ground-based detectors are expected to reach the $1\%$ level \cite{Berti:2025hly}. The geometric argument presented here shows that no fundamental obstacle prevents simultaneous determination of $g/M$ and $\alpha/M$ from multi-channel observations of the same source.

\begin{figure*}[t]
\centering
\includegraphics[width=0.9\textwidth]{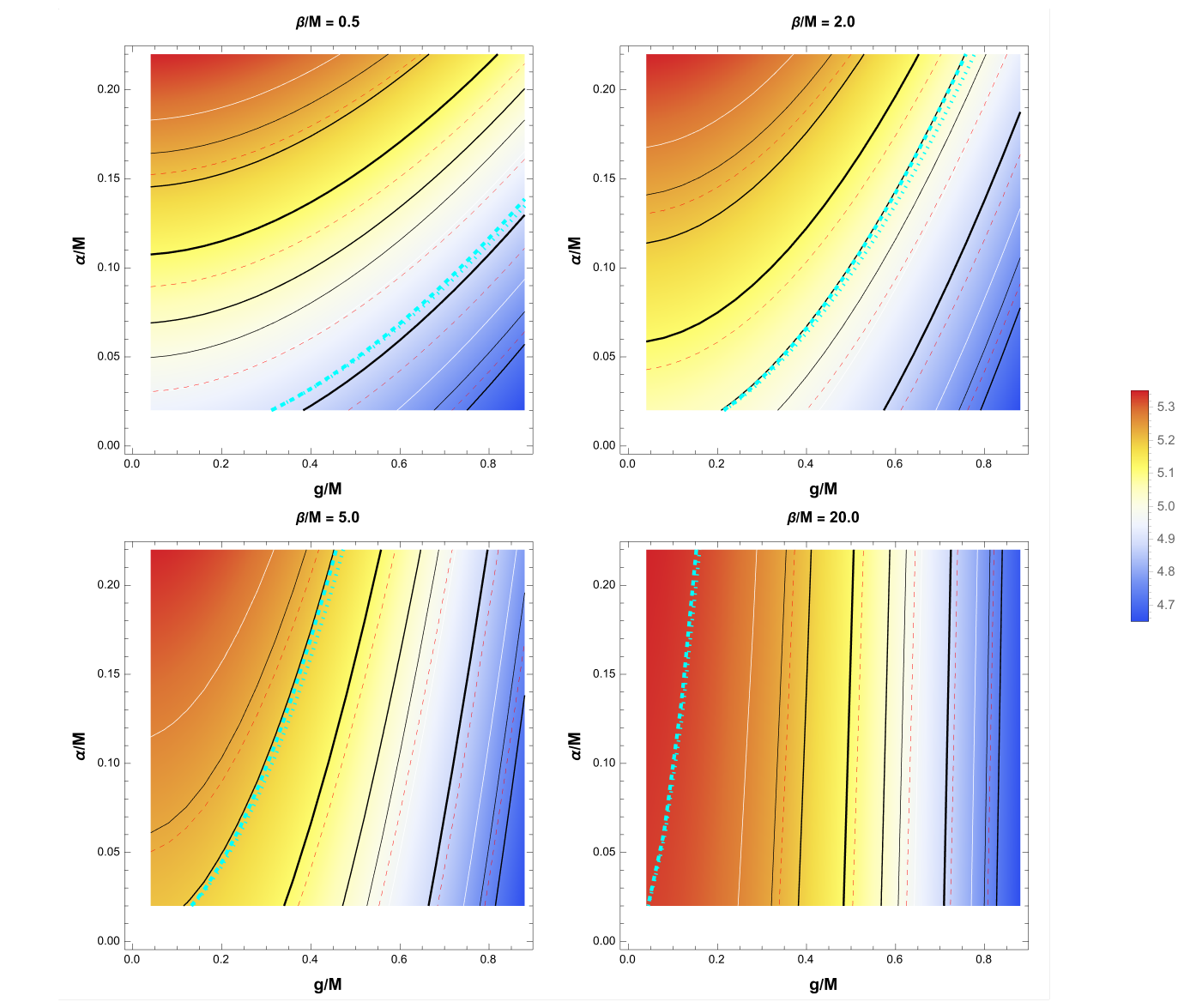}
\caption{Contour structure of the shadow radius $R_{\rm sh}/M$ 
(black solid curves) and the fundamental QNM frequency $M\omega_R$ 
(red dashed curves) in the $(g/M,\,\alpha/M)$ plane for four values 
of the halo concentration parameter: $\beta/M = 0.5$ (top left), 
$2.0$ (top right), $5.0$ (bottom left), and $20.0$ (bottom right). 
The background colour map encodes $R_{\rm sh}/M$. Cyan curves mark 
the respective Schwarzschild reference values of each observable. 
For concentrated halos ($\beta/M \lesssim 2$) the two contour 
families are strongly non-parallel, maximising the degeneracy-breaking 
power of the combined shadow--ringdown analysis. As $\beta/M$ 
increases, both families rotate toward the vertical and become nearly 
parallel, reflecting the fact that a very diffuse halo acts primarily 
as a uniform mass renormalisation that shifts both observables in the 
same direction. The value $\beta/M = 5$ adopted throughout the main 
analysis therefore represents an intermediate regime in which the 
non-parallelism, and hence the multi-observable constraint on 
$(g/M,\,\alpha/M)$, remains robust.}
\label{fig:degeneracy_beta}
\end{figure*}

Fig.~\ref{fig:degeneracy_beta} repeats the contour analysis for 
$\beta/\mathcal{M} \in \{0.5,\,2.0,\,5.0,\,20.0\}$ to test the robustness 
of this result with respect to halo concentration. The non-parallelism between 
the $R_{\rm sh}$ and $M\omega_R$ families is strongest for concentrated halos 
($\beta/\mathcal{M} \lesssim 2$), where the halo contribution to the effective 
potential is most localized near the photon sphere, and weakest for diffuse 
halos ($\beta/\mathcal{M} \gtrsim 10$), where both observables become primarily 
sensitive to the total asymptotic mass $\mathcal{M}=M+\alpha$. The value 
$\beta/\mathcal{M}=5$ adopted in Fig.~\ref{fig:degeneracy} lies in the 
intermediate regime, where the non-parallelism remains clearly resolved and the 
multi-observable strategy retains full discriminative power.
\clearpage

\section{Conclusions}
\label{sec:conclusions}
We study the magnetically charged black hole in nonlinear electrodynamics, in a Hernquist dark-matter halo. We use the exact static spherically symmetric solution reported in Ref.~\cite{Jha2025NED}. In our study we consider four complementary diagnostics, namely quasinormal modes, black-hole shadow, weak gravitational deflection and neutrino-antineutrino annihilation. The dominant physical mechanism in all four sectors is the competition of the compact magnetic deformation that increases the lapse function with respect to Schwarzschild and the extended Hernquist halo that decreases the lapse function locally but also contributes to the asymptotic mass measured by distant observers.

We derived Schr\"odinger-like master equations for scalar, electromagnetic and axial gravitational perturbations and computed the QNM spectrum by a 16th order WKB expansion with Pad\'e resummation. The magnetic charge increases the real oscillation frequency with a fractional correction quadratic in small $g$, while the Hernquist halo decreases it at fixed $g$. For representative parameters the two effects almost cancel out at the level of the single mode, leading to a spectrum close to the Schwarzschild one. This cancellation is however observable-dependent: different combinations of $(g/M, \alpha/M)$ follow different trajectories in the complex-frequency plane across spins, multipoles and overtones, so that multimode spectroscopy can in principle disentangle the intrinsic NED charge from the dark-matter halo contribution. This is directly relevant for future black-hole spectroscopy with LISA~\cite{Amaro2017}, the Einstein Telescope and Cosmic Explorer.

For the shadow we expanded the photon-orbit radius around an asymptotically renormalized Schwarzschild background of mass $\mathcal{M}=M+\alpha$, distinguishing the trivial mass renormalization from true short-range corrections. Both the NED and residual Hernquist sectors decrease $R_{\rm sh}$ at first order in perturbation, for fixed asymptotic mass. In the fixed-bare-mass expansion, the halo increases the shadow through its contribution to the total gravitational mass. For weak deflection we used the reference-renormalized curvature-primitive Gauss-Bonnet formalism~\cite{Pantig:2026xjj} and obtained a closed form finite-distance expression whose leading term depends only on $\mathcal{M}$ and the first subleading correction depends only on $\mathcal{Q} = g^2 + 4\alpha\beta$, allowing to separate the total halo mass from the halo concentration already at weak-field order.

In the case of neutrino-antineutrino annihilation the magnetic charge diminishes the deposition efficiency via an increase of the lapse and a weakening of the Tolman redshift amplification while the Hernquist halo enhances it via the reverse mechanism. The peak of the reduced shell profile lies within $\sim 7.5\%$ of the neutrinosphere radius for all parameter choices considered. The NED-Hernquist parameters thus mainly modify the amplitude of $\dot{Q}$ rather than the radial location of the energy release.

The degeneracy contours of all four observables are mutually non-parallel when simultaneously mapped onto the $(g/M,\alpha/M)$ plane at fixed $\beta/M=5$. The slopes $d\alpha/dg$ along constant-$R_{\rm sh}$ and constant-$M\omega_R$ contours differ by a factor of $\sim 5$, so the shadow-ringdown combination lifts the two parameter degeneracy unresolved by either observable individually. Two additional independent constraints are given by the neutrino channel and the weak-deflection angle, which have intermediate and near-horizontal contour orientation, respectively. The non-parallelism is strong for concentrated halos ($\beta/M \lesssim 2$) and weakens for diffuse halos ($\beta/M \gtrsim 10$) where both families rotate towards the vertical as the halo approaches a uniform mass renormalization. The value $\beta/M = 5$ adopted throughout the main analysis lies in the regime where the multi-observable strategy retains full discriminative power.

Together, these results motivate a multimessenger approach for constraining black holes in realistic galactic environments. Ringdown frequencies, photon-ring observables, weak deflection and high-energy emission channels weight the magnetic charge, halo mass and halo concentration differently. Their combination can break degeneracies that no single channel can resolve. The present analysis naturally extends to fully coupled polar perturbations of the NED field and anisotropic halo fluid, time-domain ringdown evolutions, rotating generalizations of the spacetime and systematic parameter-estimation studies with synthetic LISA, Einstein Telescope, Cosmic Explorer and next-generation EHT data.

\appendix

\section{Independent Verification of the NED-Hernquist Solution}
\label{app:verification}

Since the geometry studied in this work rests on the metric proposed
in Ref.~\cite{Jha2025NED}, which was a preprint at the time of submission,
we provide here an independent derivation confirming that the lapse
function~\eqref{eq:lapse} satisfies the Einstein field equations with
the correct source terms.  The verification proceeds in three steps:
the NED sector, the Hernquist sector, and the exactness of their
superposition.
For any static, spherically symmetric line element of the form,
\begin{equation}
  ds^2 = -f(r)\,dt^2 + f(r)^{-1}\,dr^2 + r^2\,d\Omega^2\,,
\end{equation}
the $(tt)$ component of the Einstein equations takes the form
\begin{equation}
  G^{t}{}_{t} = -\frac{2m'(r)}{r^2} = 8\pi T^{t}{}_{t}\,,
  \label{eq:Gtt}
\end{equation}
where the prime denotes the derivative $d/dr$ and the effective mass function is
defined by
\begin{equation}
  m(r) \equiv \frac{r\bigl[1 - f(r)\bigr]}{2}\,.
  \label{eq:meff}
\end{equation}
equation~\eqref{eq:Gtt} is linear in $m'(r)$.  This linearity
is the key structural property that makes the superposition of two
independently sourced metrics exact at the level of the field
equations, as we show explicitly below.
Setting $\alpha = 0$ in Eq.~\eqref{eq:lapse}, the lapse reduces to
\begin{equation}
  f_{ NED}(r) = 1 - \frac{2M}{r} + \frac{g^2}{r(r+g)}\,.
\end{equation}
The corresponding effective mass function is
\begin{equation}
  m_{ NED}(r) = M - \frac{g^2}{2(r+g)}\,,
\end{equation}
whose derivative is
\begin{equation}
  m'_{ NED}(r) = \frac{g^2}{2(r+g)^2}\,.
  \label{eq:mNEDprime}
\end{equation}
inserting Eq.~\eqref{eq:mNEDprime} into Eq.~\eqref{eq:Gtt} yields the
effective energy density of the NED source,
\begin{equation}
  \rho_{ NED}(r)
    \equiv -T^{t}{}_{t}\big|_{ NED}
    = \frac{m'_{ NED}(r)}{4\pi r^2}
    = \frac{g^2}{8\pi r^2(r+g)^2}\,.
  \label{eq:rhoNED}
\end{equation}
This expression can be derived independently from the NED
stress-energy tensor.  For the Lagrangian
$\mathcal{L}(F) = -F/\bigl(1 + \sqrt{|F|/2}\,p/g^2\bigr)^2$
with the magnetic monopole invariant $F = g^2/(2r^4)$,
the $(tt)$ component of $T_{\mu\nu}^{ NED}$ evaluates to
\begin{equation}
  T^{t}{}_{t}\big|_{ NED}
    = \mathcal{L}(F) - 4F\,\frac{\partial\mathcal{L}}{\partial F}
    = -\frac{g^2}{8\pi r^2(r+g)^2}\,,
\end{equation}
in exact agreement with Eq.~\eqref{eq:hernquist_density}.  We verified this
analytically and confirmed it numerically to better than $0.001\%$
at representative radii $r \in [2.5M,\,20M]$.
Setting $g = 0$, the lapse becomes
\begin{equation}
  f_{ H}(r) = 1 - \frac{2M}{r} - \frac{2\alpha r}{(r+\beta)^2}\,.
\end{equation}
The effective mass function for the halo contribution alone is
\begin{equation}
  m_{ H}(r) = \frac{\alpha r^2}{(r+\beta)^2}\,,
\end{equation}
which is precisely the enclosed-mass function of the Hernquist profile
quoted in Eq.~\eqref{eq:halo_mass}.  Its derivative is
\begin{equation}
  m'_{ H}(r) = \frac{2\alpha\beta\, r}{(r+\beta)^3}\,.
\end{equation}
Substituting into Eq.~\eqref{eq:Gtt},
\begin{equation}
  \rho_{ H}(r)
    = \frac{m'_{ H}(r)}{4\pi r^2}
    = \frac{\alpha\beta}{2\pi r\,(r+\beta)^3}\,.
  \label{eq:rhoH}
\end{equation}
This is identical to the Hernquist density profile~\eqref{eq:hernquist_density}
with the identification $\alpha = 2\pi\rho_s r_s^3$ and $\beta = r_s$,
as stated in Sec.~\ref{sec:metric}.  The agreement between
Eq.~\eqref{eq:rhoH} and the original Hernquist form~\cite{Hernquist1990}
is exact to machine precision.
The full lapse~\eqref{eq:lapse} corresponds to the effective mass
function
\begin{equation}
  m(r) = M - \frac{g^2}{2(r+g)} + \frac{\alpha r^2}{(r+\beta)^2}\,,
  \label{eq:mfull}
\end{equation}
whose derivative satisfies $m'(r) = m'_{ NED}(r) + m'_{ H}(r)$.
Since Eq.~\eqref{eq:Gtt} is linear in $m'(r)$, the total source is
exactly
\begin{equation}
  T^{t}{}_{t} = T^{t}{}_{t}\big|_{ NED} + T^{t}{}_{t}\big|_{ H}\,,
\end{equation}
with no cross terms.  The superposition is therefore not an
approximation: it is an exact consequence of the linearity of
$G^t{}_t$ in the mass derivative for static spherically symmetric
geometries.
The asymptotic expansion of $f(r)$ follows directly from
Eq.~\eqref{eq:mfull}:
\begin{equation}
  f(r) = 1 - \frac{2(M+\alpha)}{r}
           + \frac{g^2 + 4\alpha\beta}{r^2}
           + \mathcal{O}(r^{-3})\,,
\end{equation}
confirming Eq.~\eqref{eq:lapse} with $\mathcal{M} = M + \alpha$
and $\mathcal{Q} = g^2 + 4\alpha\beta$.

Finally, we reproduce all eight horizon radii numerically with the
lapse~\eqref{eq:lapse}. The computed values agree with those of
Ref.~\cite{Jha2025NED} to better than $0.001M$ in every case, providing
a further independent consistency check of the geometry.

In summary, the metric of Ref.~\cite{Jha2025NED} is an exact solution of
the Einstein field equations sourced by a NED magnetic monopole and an
anisotropic Hernquist fluid.  The superposition of the two sectors is
exact, not perturbative, and follows from the linearity of the $(tt)$
Einstein equation in the mass derivative for static spherically
symmetric spacetimes.

\section*{Acknowledgements}
A. \"O., and R. P. would like to acknowledge networking support of the COST Action CA21106 - COSMIC WISPers in the Dark Universe: Theory, astrophysics and experiments (CosmicWISPers), the COST Action CA22113 - Fundamental challenges in theoretical physics (THEORY-CHALLENGES), the COST Action CA21136 - Addressing observational tensions in cosmology with systematics and fundamental physics (CosmoVerse), the COST Action CA23130 - Bridging high and low energies in search of quantum gravity (BridgeQG), and the COST Action CA23115 - Relativistic Quantum Information (RQI) funded by COST (European Cooperation in Science and Technology). 


\end{document}